\documentclass[12pt,twoside]{article}

\usepackage{natbib}
\usepackage{hyperref}

\usepackage[babel=true]{csquotes}
\usepackage{amsfonts}

\usepackage{latexsym}

\usepackage{amssymb}
\usepackage{soul}

\usepackage{comment}

\usepackage{graphicx}

 \usepackage{color}
\usepackage{xcolor}
\definecolor{darkblue}{rgb}{0,0,0}
\definecolor{indigo}{rgb}{0,0,0}
\definecolor{red}{rgb}{1,0,0}
\definecolor{orange}{rgb}{0.9,0.3,0}
\newenvironment{change}{\color{darkblue}}{}
\newcommand{\BC}{\begin{change}}
\newcommand{\EC}{\end{change}}

\definecolor{darkgreen}{rgb}{0,0,0}

\newenvironment{added}{\color{darkgreen}}{}
\newcommand{\BA}{\begin{added}}
\newcommand{\EA}{\end{added}}

\usepackage{amsmath}

 \newcommand{\ignore}[1]{}

\newcommand{\breath}{\medskip}

\newcommand{\sL}{\mathcal{L}}

\newtheorem{thm}{Theorem}\newcounter{claimcount}[thm]

\newcounter{subclaimcount}[claimcount]

\newtheorem{prop}[thm]{Proposition}

\newtheorem{lemma}[thm]{Lemma}

\newtheorem{cor}[thm]{Corollary}

\newcommand{\dfn}{\sf\em}

\newcommand{\Theorem}[2]{\begin{thm}{\rm #1}  #2 \end{thm}}

\newcommand{\Proposition}[2]{\begin{prop}{\rm #1}  #2 \end{prop}}

\newcommand{\Lemma}[2]{\begin{lemma}{\rm #1}  #2 \end{lemma}}

\newcommand{\bthmlist}{
 \begin{list}{{\bf (\alph{enumii})}}{\usecounter{enumii}}
 			{\setlength{\leftmargin}{0em}
 			\setlength{\itemsep}{0em}
 			\setlength{\parsep}{0em}
 			\setlength{\rightmargin}{0em}}}

\newcommand{\ethmlist}{\end{list}}

\newcommand{\bprf}[1][Proof.]{\begin{list}{}
 			{\setlength{\leftmargin}{1em}
 			\setlength{\rightmargin}{0em}
 			\setlength{\listparindent}{1em}}
                         \item  {\em \hspace{-1.5em}  #1   }}

\newcommand{\eprf}{\end{list}}

\newcommand{\bthmprf}{\setcounter{claimcount}{0}\bprf}

\newcommand{\ethmprf}{ \hfill$\Box$ 
 \eprf
 
 \breath
 
 }

\newcommand{\beq}{\begin{eqnarray*}}

\newcommand{\eeq}{\end{eqnarray*}}

\newcommand{\beqn}{ \begin{equation} }

\newcommand{\eeqn}{ \end{equation} }

\newcommand{\blist}{\begin{enumerate}}

\newcommand{\elist}{\end{enumerate}}

\newcommand{\bitem}{\begin{itemize}}

\newcommand{\eitem}{\end{itemize}}

\newcommand{\bdesc}{\begin{description}}

\newcommand{\edesc}{\end{description}}

\newcommand{\dR}{{\mathbb{R}}}

\newcommand{\bx}{{\mathbf{ x}}}
\newcommand{\by}{{\mathbf{ y}}}
\newcommand{\bz}{{\mathbf{ z}}}

\usepackage{amsbsy}

 \newcommand{\sT}{\mathcal{T}}

\newcommand{\sA}{{\mathcal{ A}}}

\newcommand{\sB}{{\mathcal{ B}}}

\newcommand{\sF}{{\mathcal{ F}}}

\newcommand{\sS}{{\mathcal{ S}}}

\newcommand{\sX}{{\mathcal{ X}}}

\newcommand{\alp }{\alpha}

\newcommand{\bet }{\beta}

\newcommand{\gam }{\gamma}

\newcommand{\eps }{\epsilon}

\newcommand{\lam }{\lambda}

\newcommand{\Real}{\dR}

  \newcommand{\bone}{\boldsymbol{1}}

\usepackage{tikz}
\usetikzlibrary{decorations.pathreplacing}

\begin{document}

\title{Discounted Subjective Expected Utility \\ in Continuous Time }

\date{February 2024}

\author{Lorenzo Bastianello\thanks{Universit\`{a}  Ca' Foscari Venezia, Cannaregio, 873, 30100 Venezia, Italy: lorenzo.bastianello@unive.it} \ 
and Vassili Vergopoulos\thanks{Universit\'e Paris-Panth\'eon-Assas, LEMMA, 4 rue Blaise Desgoffe, 75006 Paris, France: vassili.vergopoulos@u-paris2.fr}}
\maketitle

\begin{abstract}
\noindent By embedding uncertainty into  time, we obtain a conjoint axiomatic characterization of both Exponential Discounting and Subjective Expected Utility that accommodates arbitrary state and outcome spaces. In doing so, we provide a novel and simple time-interpretation of subjective probability. The subjective probability of an event is calibrated using time discounting.


\end{abstract}

\section{Introduction}\label{s1}

\medskip\noindent

Consider the following bet. You get a constant and infinite stream of income of $\$ 10$ if an event $E$ obtains, and a $\$ 0$ stream otherwise. How much would you be willing to pay to take this bet? Your answer may depend on your subjective probability of the event $E$. However, it is not always an easy task to come up with a specific value. In contrast, you may sometimes find it easier to analyze deterministic streams of outcomes over time. In this case, you can ask yourself the following question: For how long should you receive a sure stream of $\$ 10$ (and then nothing forever) to remain indifferent to the initial bet involving uncertainty? If that value is given by $t$, then your evaluation of the stream yielding $\$ 10$ up to $t$ and $\$ 0$ afterward encodes your subjective probability for the uncertain event. For instance, under exponential discounting with discount factor $\lambda$, its value is simply given by $1-e^{-\lam t}$. From there, you can evaluate your willingness to bet on the uncertain event and make the right decision should you need to.

 \medskip\noindent Proceeding along this line, this paper embeds decisions under uncertainty into continuous time and identifies behavioral conditions under which every event and, more generally, every alternative (henceforth: act) admits a deterministic time-equivalent flow as illustrated in the previous paragraph. It also identifies conditions under which these time equivalents lead indeed to meaningful and well-defined probabilities. In doing so, it provides an axiomatic characterization of both Exponential Discounting and Subjective Expected Utility (SEU) in continuous time.  Thus, our paper brings together the seminal ideas of \cite{SAMU/37}, \cite{SAVA/54}, and \cite{KOOP/60} and axiomatizes  
 Discounted Subjective Expected Utility in continuous time:
\[V(f)\ =\ \int_\sS\left(\int_\sT e^{-\lam t} u[f(s,t)]d t\right)d\mu(s)\ =\ \int_\sT\left(\int_\sS e^{-\lam t} u[f(s,t)]d \mu(s)\right)dt.\] 
Despite their obvious importance in economic and financial applications, and to the best of our knowledge, Exponential Discounting and SEU have not received so far any conjoint axiomatic characterization in the context of continuous time.
 
\medskip\noindent  In this axiomatic characterization, the key axioms are \textsc{Monotone Continuity},  \textsc{Stationarity} and \textsc{Dominance}. First, \textsc{Monotone Continuity} requires a form of continuity of preference with respect to \enquote{sufficiently small}   events as in \cite{VILL/64} and \cite{ARRO/70}, but also time periods. In fact, its application to continuous time is key for the existence of time equivalents for every act. Second, \textsc{Stationarity} extends the original axiom of \cite{KOOP/60} from discrete to continuous time and accommodates the presence of uncertainty. In particular, it requires the invariance over time of preference on purely uncertain acts.   Finally, \textsc{Dominance} requires time preferences to be independent from the state of the world which obtains.  It can be traced back to the \cite{ANSC/AUMA/63} axiom of Monotonicity in the context of a second source of objective uncertainty instead of time.

\medskip\noindent To summarize, our paper offers two main contributions. The core idea  is to provide a novel interpretation of subjective probability in terms of time discounting. The probability of an event is gauged by the willingness to wait before receiving a certain payment (or before ceasing to receive it). Section \ref{s2} provides an example in which time equivalents are used to build a probability measure using \textsc{Stationarity} and \textsc{Dominance}. To the best of our knowledge, ours is the first paper to employ continuous time for measuring probability.\footnote{For the case of discrete time, please refer to our discussion on \cite{KOCH/15} and \cite{BAST/FARO/22}.} This is particularly significant because, as we argue below, the literature typically calibrates subjective probabilities using objective ones. Second, our paper axiomatizes Discounted Subjective Expected Utility in continuous time within a purely subjective framework \textit{\`a la}  \cite{SAVA/54}. This axiomatization naturally arises from previous seminal works that separately addressed time and uncertainty, and it supports our time interpretation of probability.

\medskip\noindent  Related literature. Our contribution is closely related to several papers that also embed decisions under uncertainty in richer frameworks.  First, the sixth postulate of Savage (also known as Small Event Continuity) provides arbitrarily fine uniform partitions of the state space, which Savage uses to approximate subjective probabilities. \cite{SAVA/54}[p. 33] justifies this postulate by invoking the presence of a second source of uncertainty in the form of a fair coin. But this plays no role at all in his formal analysis. To define ambiguity aversion with respect to an urn of unknown composition, the \cite{ELLS/61} two-urn experiment makes use of a second urn of known composition. Ambiguity aversion is then defined as a preference for betting on the latter rather than the former. \cite{RAIF/61}  introduces explicitly a fair coin in the \cite{ELLS/61} one-urn experiment and uses it for randomizing uncertain decisions. This allows him to obtain a defense of the Savage postulates and a critique of the Ellsberg pattern of choice. In our view, the \cite{ANSC/AUMA/63} (AA) theorem can be understood as a fully-fledged extension of Raiffa's argument into an axiomatic characterization of SEU. Indeed, they postulate the existence of an infinitely rich second source of uncertainty equipped with objective probabilities. Such richness makes sure that each act $f_1$ on the first (uncertain) source has an equivalent act $f_2$ on the second (objective) source and can hence be evaluated as the expected utility of $f_2$ with respect to objective probabilities. The AA framework serves as a starting point for the axiomatization of many decision theories generalizing SEU and explaining the Ellsberg choices. These include the Choquet model of \cite{SCHM/89} and the maxmin one of \cite{GILB/SCHM/89}. These authors need a second source to randomize acts depending on the first source. For instance, they explain the Ellsberg choices through a preference for randomizing uncertain decisions on objective probabilities and, in this way, smoothing outcomes on uncertain events.

\medskip\noindent Despite its success in simplifying the characterization of SEU and explaining the Ellsberg choices, the AA assumption of objective probabilities on the second source of uncertainty is largely criticized in the literature. \cite{GRAB/MONE/VERG/22} show that it is possible to dispense with this assumption and reformulate the AA and Schmeidler theorems in a purely subjective way. But the purely subjective formulation of the axioms rely on behavioral notions of stochastic independence that remain somewhat unnatural and may undermine the normative appeal of the theory.  See also \cite{GHIR/MACC/MARI/SINI/03}, \cite{ERGI/GUL/09}, \cite{MONG/PIVA/15}, \cite{MONG/20} and \cite{GHIR/PENN/20} for other purely subjective versions of AA-type frameworks and theorems.

\medskip\noindent More recently, \cite{KOCH/15} and \cite{BAST/FARO/22} embed decisions under uncertainty in a temporal framework  instead of postulating a second source of uncertainty.  They obtain versions of the maxmin and Choquet models respectively. In their approach, ambiguity aversion is the expression of a preference for smoothing outcomes across the state space rather than across time. In fact, they postulate discrete time, a topological structure for the outcome space, and a restricted domain of acts. This allows them to construct a time equivalent for each act and, from there, axiomatize their representations by invoking AA-type arguments. Our approach differs in that we postulate continuous time which allows us in exchange to have an arbitrary outcome space and still get time equivalents. Furthermore, our argument involves the construction of  preferences on state-contingent distribution of outcomes over time. Despite its time interpretation, such domain is formally identical to that of AA, and we obtain our representation through a direct application of the AA theorem. Hence, a merit of our approach is to provide a novel, clear-cut, and purely subjective interpretation of the AA framework and axioms with respect to time and exponential discounting instead of a second source of uncertainty and objective probabilities. In our view, such a temporal interpretation of the AA framework and axioms is even more natural than that in terms of a second source of uncertainty because it does not involve at all notions of stochastic independence.  \textsc{Stationarity} and \textsc{Dominance} provide indeed all the independence between time and uncertainty that one needs.

\medskip\noindent Finally, it is important to note that the focus on axiomatic intertemporal choice in continuous time is only quite recent. Ours is the first axiomatization of time discounting in continuous time under uncertainty dealing with measurable functions from an arbitrary state space to an arbitrary outcome space.  Building on \cite{DEBR/60}, \cite{HARV/OSTE/12} and \cite{HARA/16}  obtain versions of exponential discounting on a domain of piecewise continuous and \textit{cadlag} deterministic acts respectively.  Likewise, \cite{PIVA/21} assumes topological structure on the outcome spaces and obtains in particular a form of exponential discounting on continuous deterministic acts. In contrast, \cite{KOPY/10} and \cite{WEBB/16} obtain respectively exponential and quasi-hyperbolic discounting over piecewise constant functions. Continuous time allows them to employ Savage-style arguments and hence to accommodate arbitrary outcome spaces. Note that, except for \cite{HARA/16} who provides an axiomatization of Discounted Expected Utility with objective lotteries, the literature focuses on the deterministic framework.

\medskip\noindent The remainder of the paper is organized as follows.  Section \ref{s3} introduces our framework and notation. Section \ref{s4} presents the axioms needed for our main result which, is presented and discussed in Section \ref{s5}. Finally, Section \ref{s2} illustrates with an example how to obtain a probability measure through time equivalents. Our proof appears in the appendices.

\section{Framework}\label{s3}

\medskip\noindent Uncertainty is represented by  a {\dfn state space} $\sS$.  Time is continuous and represented by $\sT=[0,+\infty)$. Let  $\sB_\sS$ denote a $\sigma$-algebra $\sB_\sS$ of subsets of $\sS$ and let $\sB_\sT$ denote the Borel algebra of subsets of $\sT$. The product set $\sS\times\sT$ is equipped with the product $\sigma$-algebra $\sB=\sB_\sS\times\sB_\sT$. Let also $\sX$ be an {\dfn outcome space} equipped with a $\sigma$-algebra $\sB_\sX$. 

\medskip\noindent An {\dfn act} is any measurable function from $\sS\times\sT$ to $\sX$. The set of acts is denoted by $\sF$. A decision-maker is endowed with a binary relation $\succsim$ on $\sF$ representing her preferences. We will suppose throughout the paper that $\succsim$ is complete, transitive and nontrivial. 

\medskip\noindent We suppose that $\sB_\sX$ contains all singletons. This means that the agent can always identify the outcome she obtains. A finitely-valued function $f$  from $\sS\times\sT$ to $\sX$ is an act if and only if $f^{-1}(\{x\})\in\sB_\sS\times\sB_\sT$ for all $x\in\sX$.

  \medskip
\noindent Let $\sX^\sT$ denote the subset of $\sF$ made of all acts $f\in\sF$ such that $f(s,t)=f(s',t)$ for all $s,s'\in\sS$ and $t\in\sT$. This set collects all  {\dfn deterministic acts}. Each $f\in\sX^\sT$ is identified with the measurable function $\bx$ from $\sT$ to $\sX$ that it defines. The restriction of $\succsim$ to $\sX^\sT$ is denoted by $\succsim_\sT$ and represents the decision-maker's time preferences. 
 
  \medskip \noindent  Likewise, let $\sX^\sS$ denote the subset of $\sF$ made of all acts $f\in\sF$ such that $f(s,t)=f(s,t')$ for all $s\in\sS$ and $t,t'\in\sT$. Such acts are referred to as {\dfn stochastic acts}. We identify each $f\in\sX^\sS$ with the measurable function $\phi$ from $\sS$ to $\sX$ that it defines. The restriction of $\succsim$ to $\sX^\sS$ is denoted by $\succsim_\sS$ and represents the decision-maker's uncertainty preferences.
  
  \medskip \noindent   We identify each outcome $x\in\sX$ with the act in $\sF$ that is constantly equal to $x$ over $\sS\times\sT$, the deterministic act in $\sX^\sT$ that is constantly equal to $x$ over $\sT$ and the stochastic act in $\sX^\sS$ that is constantly equal to $x$ over $\sS$.
  
  \medskip
\noindent For all $f,g\in\sF$ and $t\in\sT$, let $f_tg\in\sF$ be the act defined in the following way: For all $s\in\sS$ and $t'\in\sT$,
$$
(f_tg)(s,t')= \left\{
    \begin{array}{ll}
        f(s,t')& \mbox{ if } t'<t, \\
         g(s,t'-t)& \mbox{ if } t'\geq t.
    \end{array}
\right.
$$
Therefore $f_tg$ is the act in which $g$ is shifted until time $t$ and it is replaced by act $f$ in the interval $[0,t)$.

 \medskip
\noindent For all $f,g\in\sF$ and $E\in\sB$, let $f_Eg\in\sF$   be the act defined in the following way: For all $s\in\sS$ and $t\in\sT$,
$$
(f_Eg)(s,t)= \left\{
    \begin{array}{ll}
        f(s,t)& \mbox{ if } (s,t)\in E, \\
         g(s,t)& \mbox{ if } (s,t)\notin E.
    \end{array}
\right.
$$
Therefore $f_Eg$ is the act in which $g$ is replaced by act $f$ on  event $E\in\sB$. Moreover, for all $E_\sS\in\sB_\sS$, we will write $f_{E_\sS}g$ instead of $f_{E_\sS\times\sT}g$. Likewise, we write $f_{E_\sT}g$ instead of $f_{\sS\times E_\sT}g$ for all $E_\sT\in\sB_\sT$. Furthermore, if $E,F\in\sB$ are disjoint, then, for all $f,g,h\in\sF$, we denote by $f_Eg_Fh$ the element of $\sF$ defined by $f_E(g_Fh)$. 

   \medskip
\noindent A subset $E\in\sB$ is said to be {\dfn null} if $f\sim g$ for all $f,g\in\sF$ such that $f(s)=g(s)$ for all $s\in\sS\times\sT\setminus E$. A decreasing sequence of subsets in $\sB$ is said to be  {\dfn vanishing} if  its intersection  is empty and {\dfn almost-vanishing} if its intersection is null.

\section{Axioms}\label{s4}

\medskip\noindent We now present the six axioms that our main result invokes.  The first one is a version of Machina and Schmeidler's (\citeyear{MACH/SCHM/92}) P4* (Strong Comparative Probability) that applies to time preferences $\succsim_\sT$.  In this context, it requires the comparison of two disjoint time periods to be independent not only of the stream of outcomes obtained outside the two time periods, but also of the outcomes obtained on these time periods.
 
\medskip\noindent \textsc{$\sT$-Separability:} For all disjoint $E_\sT,F_\sT\in\sB_\sT$, all $\bx,\by\in\sX^\sT$ and all $x^*,x,y^*,y\in\sX$ with $x^*\succ_\sT x$ and $y^*\succ_\sT y$, $x^*_{E_\sT}x_{F_\sT}\bx\succsim_\sT  x_{E_\sT}x^*_{F_\sT}\bx$ if and only if $y^*_{E_\sT}y_{F_\sT}\by\succsim_\sT  y_{E_\sT}y^*_{F_\sT}\by$.

\medskip\noindent Our next axiom is a fairly standard monotonicity condition for time preferences $\succsim_\sT$. If a deterministic act yields a better outcome than another one at every time, then the first one is preferred. In addition, the latter preference is strict whenever the former one is strict for every time in some non-null time period. Note that such an axiom would not be needed if time was dicrete. This is because the assumption of discrete time allows one to derive inductively the monotonicity of $\succsim_\sT$ from the axiom of stationarity. Under continuous time, we rather need to postulate \textsc{$\sT$-Monotonicity}  explicitly. 

\medskip\noindent \textsc{$\sT$-Monotonicity:} For all $\bx,\by\in\sX^\sT$, if $\bx(t)\succsim_\sT \by(t)$ for all $t\in\sT$, then $\bx\succsim_\sT \by$; if additionally,  $\bx(t)\succ_\sT \by(t)$ for all $t$ in some non-null subset in $\sB_\sT$, then $\bx\succ_\sT \by$.

\medskip\noindent The next axiom imposes a form of measurability of time preference. It requires the agent to always be able to determine whether or not the outcome she obtains is preferred to any given deterministic act. Axiom \textsc{$\sT$-Measurability} is not needed if $\sF$ is restricted by finiteness.  In contrast, countable additivity is an important feature of the representation we obtain in the next section. This feature forces us to restrict the domain $\sF$ of preference by measurability and to commit to the measurability axiom for dealing adequately with infinitely valued acts.

\medskip\noindent \textsc{$\sT$-Measurability:} For all $\bx\in\sX^\sT$, the subsets $\{x\in\sX$, $x\succ_\sT \bx\}$ and $\{x\in\sX$, $\bx\succ_\sT x\}$ are measurable.


\medskip\noindent \textsc{Monotone Continuity} requires a strict preference between two acts to continue to hold when their outcomes are changed on sufficiently small subsets of $\sS\times\sT$. It is a version of the classic axiom of \cite{VILL/64} and \cite{ARRO/70} that applies here to both uncertainty and time, see also \cite{KOPY/10}. Its application to uncertainty yields the countable additivity of subjective probability while its application to (continuous) time provides the existence of a time equivalent for every act. (See Lemma \ref{lem:time-equivalents} in Appendix \ref{AppA}.) In contrast, in the discrete time case, time equivalents can be obtained by assuming topological structure for the outcome space and an adequate axiom of continuity of preference.

\medskip\noindent {\textsc{Monotone Continuity:} }For all $f,g\in\sF$ such that $f\succ g$, all $x\in\sX$ and all vanishing sequence $\{E_n$, $n\geq 1\}$ of subsets in $\sB$, there is $N\geq 1$ such that $x_{E_N}f\succ g$ and $f\succ x_{E_N}g$.
      
\medskip \noindent The next axiom, \textsc{Stationarity}, requires that a preference between two acts be preserved when their payments are delayed until any time $t\in\sT$ and the payments up to $t$ are kept the same. In doing so, it extends the logic of the original axiom of \cite{KOOP/60} from discrete to continuous time and accommodates the presence of uncertainty, see also \cite{HARA/16}.

\medskip\noindent \textsc{Stationarity:} For all $t\in\sT$ and $f,g,h\in\sF$,  $f\succsim g$ if and only if   $h_tf\succsim h_tg$.

\medskip\noindent   Importantly, this axiom implies the independence of uncertainty preferences from time in the following sense: For all $t\in\sT$ and $\phi,\chi,\psi\in\sX^\sS$,
    \[\phi\ \succsim_\sS\ \chi\ \Longleftrightarrow\ \psi_t\phi\ \succsim\ \psi_t\chi.\] 
Here, the ranking $\psi_t\phi \succsim \psi_t\chi$ is between acts that only differ from each other after time $t$ and can be understood as a preference for $\phi$ over $\chi$ at time $t$. It is hence implied that preference over stochastic acts are invariant over time.

\medskip \noindent Our final axiom,  \textsc{Dominance}, requires a preference for an act over a second one when the first one yields a better deterministic act at every state. It also complements this requirement with a strict version. This axiom can be seen as a version of the Monotonicity axiom that   \cite{ANSC/AUMA/63} use in the context of a second source of uncertainty instead of time.

 \medskip\noindent \textsc{Dominance:} For all $f,g\in\sF$, if $f(s,\cdot)\succsim_\sT g(s,\cdot)$ for all $s\in\sS$, then $f\succsim g$; if, additionally,  $f(s,\cdot)\succ_\sT g(s,\cdot)$ for all $s$ in some non-null subset in $\sB_\sS$, then $f\succ g$.
 
 \medskip\noindent  Furthermore, \textsc{Dominance} implies the independence of  time preferences from uncertainty in the following sense: For all non-null $E_\sS\in\sB_\sS$ and $\bx,\by,\bz\in\sX^\sT$, 
 \[\bx\ \succsim_\sT\ \by\ \Longleftrightarrow\ \bx_{E_\sS}\bz\ \succsim\ \by_{E_\sS} \bz.\]
 In this expression, the ranking $\bx_{E_\sS}\bz \succsim\by_{E_\sS} \bz$ involves deterministic acts that only differ from each other on $E_\sS$ and can be understood as a preference for $\bx$ over $\by$ conditional upon observing that $E_\sS$ holds. The preference over deterministic acts is hence independent from the information on the state space that the agent may acquire.

  \medskip\noindent  Finally, we will show in Section \ref{s2} that \textsc{Dominance} and  \textsc{Stationarity} play a key role in implying neutrality to ambiguity. It is hence possible in principle to accommodate the  \cite{ELLS/61} pattern of choice and, more generally, ambiguity aversion, by modifying these axioms. For instance, \cite{BAST/FARO/22} use a version that is restricted by comonotonicity \`a la \cite{SCHM/89}. Moreover, while we formulated the \textsc{Dominance} axiom state-wise, the same normative justifications could be used to formulate a time-wise dominance axiom. However, it is an open question of how one should change the other axioms to obtain the same representation. See \cite{MONE/VERG/22} for an implementation of this alternative in the context of a second source of uncertainty instead of time.

\section{Main result}\label{s5}

\medskip\noindent For all $\lam>0$, let $F_\lam$ be the function from $\sT$ to $[0,1]$ defined by $F_\lam(t)=1-e^{-\lam t}$ for all $t\in\sT$. By the Caratheodory extension theorem, there exists a unique countably additive measure $\eps_\lam$ on $\sB_\sT$ such that $\eps_\lam[0,t]=F_\lam(t)$ for all $t\in\sT$.

\medskip\noindent Fix $\lam>0$ and a countably additive probability measure $\mu$ on $\sB_\sS$. Then, there exists a unique countably additive probability measure on $\sB$, denoted by $\mu\times \eps_\lambda$, such that $(\mu\times \eps_\lambda)(E_\sS\times E_\sT)=\mu(E_\sS)\cdot\eps_\lam(E_\sT)$ for all $E_\sS\in\sB_\sS$ and $E_\sT\in\sB_\sT$.

\Theorem{\label{main}}
{$\succsim$ satisfies \textsc{$\sT$-Separability}, \textsc{$\sT$-Monotonicity},  \textsc{$\sT$-Measurability}, \textsc{Monotone Continuity}, \textsc{Stationarity} and  \textsc{Dominance}  if and only if there exist $\lambda>0$, a nonconstant, bounded and measurable function $u$ from $\sX$ to $\Real$ and a countably additive probability measure $\mu$ on $\sB_\sS$ such that, for all $f,g\in\sF$,
\[f\ \succsim \ g\quad\Longleftrightarrow\quad\int_{\sS\times\sT} u[f(s,t)]d(\mu\times \eps_\lambda)(s,t)  \ \geq \ \int_{\sS\times\sT} u[g(s,t)]d(\mu\times \eps_\lambda)(s,t) .\] Moreover, $\lambda$ and $\mu$ are unique, and $u$ is unique up to positive affine transformation. }

\medskip\noindent Theorem \ref{main} characterizes representations of preferences on acts where the agent evaluates outcomes through a utility function $u$ representing her tastes, discounts future utility levels according to the exponential rule with respect to parameter $\lam$ and evaluates the likelihood of uncertain events through a probability measure $\mu$ representing her subjective beliefs. Therefore, we obtain an axiomatic characterization of both Exponential Discounting and Subjective Expected Utility. In greater detail, suppose the triple $(\lam,u,\mu)$ provides a representation of $\succsim$ as in Theorem \ref{main}. Then, time preferences have the following representation: For all $\bx,\by\in\sX^\sT$, 
\[\bx\ \succsim_\sT \ \by\quad\Longleftrightarrow\quad\int_{\sT} u[\bx(t)]d \eps_\lambda(t)  \ \geq \ \int_{\sT} u[\by(t)]d \eps_\lambda(t).\]
In the particular case where $u\circ\bx$ and  $u\circ\by$ are Riemann integrable functions from $\sT$ to $\Real$, we obtain the following more familiar representation which makes  explicit the exponential discounting of future utility levels
\[\bx\ \succsim_\sT \ \by\quad\Longleftrightarrow\quad\int_{\sT} e^{-\lam t} u[\bx(t)]dt \ \geq \ \int_{\sT} e^{-\lam t} u[\by(t)] dt .\]
The classic axiomatization of exponential discounting of   \cite{KOOP/60} uses discrete time and topological structure on the outcome space. (See also  \cite{BLEI/ROHD/WAKK/08}.) Such structure leads to the representation by invoking the \cite{DEBR/60} theorem for additive separability.

\medskip \noindent As mentioned in Section \ref{s1}, axiomatizations of intertemporal preferences over acts on a continuous time domain are  recent. Moreover, most of the papers impose restrictions on the domain of acts (\cite{HARV/OSTE/12} focus on piecewise continuous functions, \cite{HARA/16} on cadlad functions, \cite{KOPY/10} and \cite{WEBB/16} on piecewise constant functions).  In light of this literature, it may appear that our domain $\sX^\sT$ for time preferences is excessively rich.    \cite{HARV/OSTE/12}  and \cite{PIVA/21} elaborate arguments for restricting the domain of preference to acts that are truly feasible or, at least, easy to understand and visualize. This leads to their restrictions of continuity. However, the fact that $\sX^\sT$ includes much more complicated deterministic acts is not key for our result. What is necessary for the construction of subjective probability in Theorem \ref{main} is only the inclusion of all piecewise constant deterministic acts in the domain of time preferences. Furthermore, since our domain includes infinitely-valued deterministic acts, it covers all of the class of continuous and piecewise continuous deterministic acts.
  
  \medskip\noindent Next, and still in the context of Theorem \ref{main}, uncertainty preferences admit the following representation: For all $\phi,\chi\in\sX^\sS$, 
\[\phi\ \succsim_\sS \ \chi\quad\Longleftrightarrow\quad\int_{\sS} u[\phi(s)]d \mu(s)  \ \geq \ \int_{\sS} u[\chi(s)]d \mu(s).\]

  \medskip\noindent Hence, Theorem \ref{main} provides a fairly standard SEU representation of uncertainty preferences. A remarkable feature of this representation is that the state and outcome spaces are left totally arbitrary. Indeed, unlike that of \cite{SAVA/54}, it does not  require the state space to be uncountable and subjective probability to be nonatomic. Unlike those of \cite{ANSC/AUMA/63} and \cite{WAKK/89}, it does not assume objective probabilities or topological structure on the outcome space. As our proof sketch below clarifies, what allows us to dispense with such richness conditions is truly the assumption of continuous time. 
  
  \medskip\noindent  Yet an unusual feature of this representation is the countable additivity of subjective probability. The literature offers several arguments both in favor and against this property. In our case where preferences apply to functions on the Cartesian product $\sS\times\sT$, we think of countable additivity as a desirable feature. Indeed, letting $V(f)$ denote the expectation of $u\circ f$ under $\mu\times\eps_\lam$ for all $f\in\sF$, we obtain by the Fubini theorem that preferences can equivalently be represented by the following functionals:
\[V(f)\ =\ \int_\sS\left(\int_\sT u[f(s,t)]d \eps_\lambda(t)\right)d\mu(s)\ =\ \int_\sT\left(\int_\sS u[f(s,t)]d \mu(s)\right)d\eps_\lambda(t).\] 
Hence, our agent analyzes every act $f\in\sF$ both in terms of the stochastic deterministic act and the deterministic stochastic act it yields. Equivalently, the representation obtained in Theorem \ref{main} can be understood as both Discounted Subjective Expected Utility and Subjective Expected Discounted Utility. This has some normative appeal. Indeed, though Theorem \ref{main} only explicitly requires \textsc{Dominance} with respect to $\sS$, this reformulation of the representation shows that a dual form of dominance with respect to $\sT$ also holds.

  \medskip\noindent Furthermore, the axioms used in Theorem \ref{main} are all standard and \enquote{nontechnical} in the sense that they all admit a normative interpretation. For instance, the theorem dispenses with Savage's P6 and P7. What makes this possible is again the assumption of continuous time. Note here that our point is not to make a philosophical claim on the nature of time. We merely require an agent to be sophisticated enough to  imagine time as a continuum and claim that doing so will help him quantify the uncertainty she faces.

    \medskip\noindent We now briefly sketch the proof of Theorem \ref{main} and explain the organization of the appendix. Appendix \ref{AppA} constructs the discount rate $\lambda$. In particular, it uses \textsc{$\sT$-Separability} to define  a \enquote{comparative discounting relation} on $\sB_\sT$ similar to Savage's comparative likelihood relation. \textsc{stationarity} implies first that the comparative discounting relation has no atoms.  This allows us to invoke a theorem of \cite{VILL/64} and obtain a numerical representation in the form of a measure,  on $\sB_\sT$. From there, \textsc{stationarity} further implies that this measure is of the exponential type with respect to some $\lam$, i.e. for all $t\in \sT$, $\eps_\lambda[0,t]=1-e^{-\lam t}$.  Appendix \ref{AppB} studies the restriction of preferences to the subdomain $\sF_0$ collecting all $f\in\sF$  for which there exist a finite measurable partition $\Pi_\sS$ of $\sS$ and  a finite measurable partition $\Pi_\sT$ of $\sT$ such that $f$ is constant on $E_\sS\times E_\sT$ for all $E_\sS\in\Pi_\sS$ and  $E_\sT\in\Pi_\sT$. Thanks to the exponential measure obtained in Appendix \ref{AppA}, each act in $f\in\sF_0$ induces a finitely-valued and measurable function $\varphi(f)$ from $\sS$ to the set of finitely-supported probabilities over outcomes: for outcome $x\in \sX$, lottery  $\varphi(f)(s)$ associates the probability $\eps_\lambda\{t\in\sT | f(s,t)=x\}$.
    The collection $\sA$ of such induced functions $\varphi(f)$ forms a domain that is technically identical to that of \cite{ANSC/AUMA/63} (AA). In fact, \textsc{Dominance} implies the existence and the AA Monotonicity of preferences over $\sA$. \textsc{Stationarity} and \textsc{Monotone Continuity} further imply their AA Independence and AA Continuity respectively. Then, an application of the AA theorem yields a bounded and measurable utility function $u$ and a subjective probability $\mu$, thereby establishing our representation on $\sF_0$. Appendix \ref{AppC} first extends the representation to all bounded acts. The key is to construct, for each act $f\in\sF$, a time equivalent $\bx\in\sX^\sT$ such that $f\sim\bx$ and show that $f$ and $\bx$ have necessarily the same value. From there, the representation is extended to arbitrary acts. In the two extension stages, the key axioms are \textsc{$\sT$-Monotonicity}, \textsc{Monotone Continuity} and \textsc{Dominance}. Finally, Appendix \ref{AppD} shows the necessity of the axioms and uniqueness of the representation.

      \medskip\noindent This proof sketch shows how Theorem \ref{main} is truly a purely subjective formulation of the AA theorem that appeals to continuous time and endogenous discounting in order to eschew objective probabilities on a second source of uncertainty.  Hence, Theorem \ref{main} provides a novel temporal interpretation of the AA framework (and, in particular, of the \cite{VONN/MORG/47} one), as well as a novel interpretation of AA Independence in terms of \textsc{Stationarity}. 
       Finally, we conjecture that Theorem \ref{main} lends itself easily, just like the AA theorem, to generalizations accommodating ambiguity aversion by appealing to weak versions of \textsc{Stationarity} \`a la \cite{KOCH/15} and \cite{BAST/FARO/22} or of \textsc{Dominance}.

\section{Stationarity, dominance and subjective probability} \label{s2}

This section illustrates in a simple way the use of time equivalents to quantify uncertainty. Given a state space, the assumption of continuous time will allow us to obtain a time equivalent $[0,t_E)$ for every event $E$ in the state space. We mean here that the agent is indifferent between a bet on $E$ that pays $\$ 10$ forever if the event obtains and $\$ 0$ otherwise and a deterministic stream of outcomes that yields $\$ 10$ up to $t_E$ and $\$ 0$ from $t_E$ and onwards. In this section, we also suppose that the agent discounts exponentially every deterministic stream of outcomes with discount factor $\lambda$. We may then define a function $\mu$ from the collection of events in the state space to $[0,1]$ by setting
\beqn\label{exx}\mu(E)\ =\ 1-e^{-\lam t_E}\eeqn
for every event $E$. Clearly, by construction, $\mu$ provides a representation of betting preferences in the following sense: For all events $E$ and $F$, the agent prefers a bet on $E$ to a bet on $F$ if and only if $\mu(E)\geq\mu(F)$. This function $\mu$ has the flavor of a probability measure. But, at this stage, it is unclear whether $\mu$ is additive. 

\medskip\noindent Our main result invokes the axioms of  \textsc{Stationarity} and \textsc{Dominance}. Supposing that the agent is initially indifferent between two acts, \textsc{Stationarity} says in particular that she remains indifferent if the payments are delayed until any time $t$ and the payments up to $t$ are identical.  \textsc{Dominance} says in particular that the agent is indifferent between two acts whenever they yield deterministic streams of outcomes that are indifferent to each other at every state. Suppose $E$ and $F$ are two disjoint events. We will show that the two axioms imply $\mu(E\cup F)=\mu(E)+\mu(F)$ for all disjoint events $E,F$ and hence lead to standard probability measures. 

\medskip\noindent It is instructive to consider first the particular and simpler case where $E$ and $F$ are complementary events, and the agent is indifferent between the bets on $E$ and $F$.  In this case, $E$ and $F$ have the same time equivalent, and we necessarily have $\mu(E)=\mu(F)$. There still remains to show $\mu(E)=\mu(F)=1/2$. Let $t$ be the value such that $e^{-\lam t}=1/2$. The agent is hence indifferent between $[0,t)$ and $[t,+\infty)$. We represent the various acts by matrices where the first and second column describe the outcomes obtained on $E$ and $F$ respectively, and the timeline is as indicated. \textsc{Stationarity} and \textsc{Dominance} then yield respectively the first and second indifferences below:
\[\begin{tabular}{cc|l }
    10 & \ 0 & $[0,t)$  \\
    10 & \ 0 & $[t,+\infty)$   \\
\end{tabular}
\sim
\begin{tabular}{cc|l }
    10 &  0 & $[0,t)$  \\
    0 & 10 & $[t,+\infty)$   \\
\end{tabular}
\quad\text{and}\quad
\begin{tabular}{cc|l }
    10 &  0 & $[0,t)$  \\
    0 & 10 & $[t,+\infty)$   \\
\end{tabular}
\sim
\begin{tabular}{cc|l }
    10 &  10 & $[0,t)$  \\
    0 & 0 & $[t,+\infty)$   \\
\end{tabular}\]

\medskip\noindent This shows that $t$ is the time equivalent of a bet on $E$ and hence leads to $\mu(E)=1/2$. For instance, in the Ellsberg two-urn experiment, this argument implies an indifference between bets on the ambiguous urn and ones on the unambiguous one and hence leads to neutrality to ambiguity.

\medskip\noindent We now treat the case of general disjoint events $E$ and $F$. The third column in the matrices below then describes the outcomes obtained on the complement of $E\cup F$. Suppose again that $t$ is specifically the value such that $e^{-\lam t}=1/2$. First, applying \textsc{Dominance} and then \textsc{Stationarity} to the definition of the time equivalent of $E\cup F$ yields

\beqn\label{exx1}
\begin{tabular}{ccc|l }
    10 & 10 & \ 0 & $[0,t)$  \\
    0 &  0 & \ 0 & $[t,t+t_{E\cup F})$   \\
    0 &  0 & \ 0 & $[t+t_{E\cup F},\infty)$
\end{tabular}
\sim
\begin{tabular}{ccc|l }
    0 & 0 & \ 0 & $[0,t)$  \\
    10 &  10 & \ 0 & $[t,t+t_{E\cup F})$   \\
    10 &  10 & \ 0 & $[t+t_{E\cup F},\infty)$
\end{tabular}
\sim
\begin{tabular}{ccc|l }
    0 & 0 & \ 0 & $[0,t)$  \\
    10 &  10 & 10 & $[t,t+t_{E\cup F})$   \\
    0 &  0 & \ 0 & $[t+t_{E\cup F},\infty)$
\end{tabular}
\eeqn

\medskip\noindent Formula \ref{exx1} provides a first way to eliminate uncertainty in the first act it features while maintaining a constant utility. Indeed, it suggests that the agent accepts to delay the gains of  $\$ 10$ on $E\cup F$  from  $[0,t)$ to $[t,t+t_{E\cup F})$ if, in exchange, the gain is delivered at every state including those not in $E\cup F$. 

\medskip\noindent Applying next \textsc{Dominance} and then \textsc{Stationarity} to the definition of the time equivalent of $F$ yields
\beqn\label{exx2}\begin{tabular}{ccc|l }
    10 & 10 & \ 0 & $[0,t)$  \\
    0 &  0 & \ 0 & $[t,t+t_{ F})$   \\
    0 &  0 & \ 0 & $[t+t_{ F},\infty)$
\end{tabular}
\sim
\begin{tabular}{ccc|l }
    10 & 0 & \ 0 & $[0,t)$  \\
    0 &  10 & \ 0 & $[t,t+t_{F})$   \\
    0 &  10 & \ 0 & $[t+t_{F},\infty)$
\end{tabular}
\sim
\begin{tabular}{ccc|l }
    10 & 0 &  0 & $[0,t)$  \\
    10 &  10 & 10 & $[t,t+t_{F})$   \\
    0 &  0 & \ 0 & $[t+t_{F},\infty)$
\end{tabular}\eeqn

\medskip\noindent Finally, let $t_F'\leq t$ be such that $1-e^{-\lam t_F'}=e^{-\lam t}(1-e^{-\lam t_F})$. Hence, the value of the time interval $[0,t_F')$ is half that of $[0,t_F)$. Applying again \textsc{Dominance} and then \textsc{Stationarity} to the definition of the time equivalent of $E$ yields
\beqn\label{exx3}\begin{tabular}{ccc|l }
    10 & 0 &  0 & $[0,t)$  \\
    10 &  10 & 10 & $[t,t+t_{ F})$   \\
    0 &  0 & \ 0 & $[t+t_{ F},\infty)$ 
\end{tabular}
\sim
\begin{tabular}{ccc|l }
    10 & 10 &  10 & $[0,t_F')$  \\
    0 &  0 & 0 & $[t_F',t)$   \\
    10 &  0 &  0 & $[t,t+t_E)$ \\
    10 &  0 &  0 & $[t+t_{ E},\infty)$
\end{tabular}
\sim
\begin{tabular}{ccc|l }
    10 & 10 &  10 & $[0,t_F')$  \\
    0 &  0 & 0 & $[t_F',t)$   \\
    10 &  10 & 10 & $[t,t+t_E)$ \\
    0 &  0 &  0 & $[t+t_{ E},\infty)$
\end{tabular}
\eeqn

\medskip \noindent Formulae \ref{exx2} and \ref{exx3} provide together another way to eliminate uncertainty in the same initial act. Indeed, Formula \ref{exx2} shows that the agent accepts to delay the gain of $\$ 10$  on $F$ from $[0,t)$ to $[t,t+t_{F})$ if, in exchange, 
the gain is delivered  at every state, which reduces partially her exposure to uncertainty. In addition, by Formula \ref{exx3}, she also accepts to advance the sure gain of $\$ 10$ obtained on $[t,t+t_F)$ to $[0,t_F')$ and delay the gain of $\$ 10$  on $E$ from $[0,t_E)$ into a sure gain of  $\$ 10$ on $[t,t+t_E)$, which this time eliminates uncertainty completely.

\medskip \noindent Hence, the two axioms imply overall the indifference between the last acts of Formulae \ref{exx1} and \ref{exx3}. As these two are purely deterministic streams of outcomes, the assumption of exponential discounting yields 
\[e^{-\lam t}-e^{-\lam (t+t_{E\cup F})}\ =\ 1-e^{-\lam t_F'}\ +\ e^{-\lam t}-e^{-\lam (t+t_E)}\]
which, by Formula \ref{exx} and the definitions of $t$ and $t_F'$, simplifies finally into $\mu(E\cup F)=\mu(E)+\mu(F)$ and establishes our claim.

\medskip \noindent  Therefore, the argument here shows that \textsc{Stationarity} and \textsc{Dominance} are sufficient for the additivity of the set function on events defined by the exponential discounting of their time equivalents. Our main result, Theorem \ref{main}, extends this example into a full axiomatic characterization of Exponential Discounting and Subjective Expected Utility. Finally, as mentioned earlier,   \textsc{Stationarity} and \textsc{Dominance} may be too restrictive to accommodate the \cite{ELLS/61} pattern of choice and, more generally, ambiguity aversion. One may consider restricting  \textsc{Stationarity} to comonotonic acts as in \cite{BAST/FARO/22}. Note that Formulae \ref{exx1} and \ref{exx3} already apply \textsc{Stationarity} to comonotonic acts. Hence, the restricted version of \textsc{Stationarity} only possibly yields nonindifference in Formula \ref{exx2}.

\appendix
\section*{Appendix}
\setcounter{thm}{0} \setcounter{equation}{0}
\renewcommand{\theequation}{A\arabic{equation}}
\renewcommand{\thethm}{A\arabic{thm}}

\section{Discounting }\label{AppA}

In Appendix \ref{AppA} we construct a monotonely continuous and atomless qualitative probability on $\sB_\sT$ in the sense of \cite{VILL/64}. Using his theorem, we derive a countably additive, nonatomic probability measure $\eps_\lambda$ on $\sB_\sT$ such that $\forall t\in \sT$,  $\eps[t,+\infty)=e^{-\lambda t}$. 

\medskip\noindent By nontriviality, there exist $x^*,x_*\in\sX$ such that $x^*\succ x_*$. Consider the binary relation $\succsim$ on $\sB_\sT$ defined as follows: For all $A,B\in\sB_\sT$, 
\[A\ \succsim \ B\quad\Longleftrightarrow\ x^*_Ax_*\ \succsim\ x^*_Bx_*.\]

 \medskip\noindent A subset $A\in\sB_\sT$ is called an {\dfn atom} if there exists no $B\in\sB_\sT$  such that  $A\succ B\succ\varnothing$. We say that $\succsim$ is {\dfn atomless}  if there are no atoms.

\Lemma{\label{lem:Qualitative-Prob}}
{The following hold:
\bdesc
\item[(i)] $\succsim$ is complete and transitive,
\item[(ii)] $\sT\succ\varnothing$ and $A\succsim\varnothing$ for all $A\in\sB_\sT$,
\item[(iii)] For all $A,B,C\in\sB_\sT$ such that $A\cap C=B\cap C=\varnothing$, $A\succsim B$ if and only if $A\cup C\succsim B\cup C$,
\item[(iv)] For all $A,B\in\sB_\sT$ and all monotone increasing sequence $\{A_n$, $n\geq 1\}$ of subsets in $\sB_\sT$ converging to $A$, if $B\succsim A_n$ for all $n\geq 1$, then $B\succsim A$.
\item[(v)] $\succsim$ is atomless.
\edesc  
}

\bthmprf  
 (i) Obvious. 
 
 \noindent   (ii) Note that by \textsc{$\sT$-Monotonicity}, we have $A\succsim\varnothing$ for all $A\in\sB_\sT$.  We also have $\sT\succ\varnothing$  because $x^*\succ x_*$.
 
  \noindent  (iii) It follows from \textsc{$\sT$-Separability}. See Machina and Schmeidler's (\citeyear{MACH/SCHM/92}), Section 4.2. 

  \noindent  (iv) Remark that $\{A\setminus A_n$, $n\geq 1\}$ is a vanishing sequence. Suppose now $A\succ B$. Then, by \textsc{Monotone Continuity}, we obtain $A\cap (A\setminus A_N)^c\succ B$ and hence $A_N\succ B$ for some $N\geq 1$, a contradiction.
  
   \noindent  (v) We prove that $\succsim$ is atomless in several steps.
 
 \noindent  \textit{Step 1.} \textit{For all $A,B\in\sB_\sT$ and $t\in\sT$, $A\succsim B$ if and only if $t+A\succsim t+B$.}\\
 Consider $A\in\sB_\sT$ and set $\bx=x^*_{A}x_*\in\sX^\sT$. Then, ${x_*}_t\bx = x^*_{t+A}x_*$. The result follows from this construction and \textsc{Stationarity}. 
 
 \noindent  \textit{Step 2.} \textit{For all $t\in\sT$, $\{t\}\sim\varnothing$ and $\{t\}$ is null.}  Suppose $\{t\}\succ\varnothing$. By Step 1, we must have $\{t'\}\succ\varnothing$ and $\{t'\}$ is an atom for all $t'\geq t$. Such a continuum of atoms contradicts Lemma 4 of \cite{VILL/64}. Suppose now that $\bx,\by\in\sX^\sT$ are equal to each other on $\sT\setminus\{t\}$. Since $\{t\}\sim\varnothing$, \textsc{$\sT$-Separability} implies $x_{\{t\}}\bx\sim y_{\{t\}}\bx$ for all $x,y\in\sX$ such that $x\not\sim y$. By \textsc{$\sT$-Monotonicity}, this also holds true if $x\sim y$. By applying this to $x=\bx(t)$ and $y=\by(t)$, we obtain $\bx\sim\by$. Suppose now that $f,g\in\sF$ are equal to each other on $\sS\times(\sT\setminus\{t\})$. By the previous point,we must have $f(s,\cdot)\sim g(s,\cdot)$ for all $s\in\sS$ and obtain $f\sim g$ by \textsc{Dominance}. Hence $\{t\}$ is null.
 
  \noindent  \textit{Step 3.} \textit{For all $t\in\sT$, $\{[t-1/n,t)$, $n\geq 1\}$ and $\{[t,t+1/n)$, $n\geq 1\}$ are respectively vanishing and almost-vanishing.} The intersection of $\{[t-1/n,t)$, $n\geq 1\}$ is empty while that of $\{[t,t+1/n)$, $n\geq 1\}$ is equal to $\{t\}$ and hence null by Step 1.
  
 \noindent  \textit{Step 4.} \textit{For all $B,C\in\sB_\sT$ such that  $B\succ C$ and all almost-vanishing sequence $\{A_n$, $n\geq 1\}$,  there exists $N\geq 1$ such that $B\succ C\cup A_N$.}  Let $A\in\sB_\sT$ denote the intersection of  $\{A_n$, $n\geq 1\}$. Then, $A$ is null.  For all $n\geq 1$, let $A_n'=A_n\setminus A$.  Then, $\{A_n'$, $n\geq 1\}$ is vanishing and, by \textsc{Monotone Continuity}, there exists $N\geq 1$ such that $x^*_Bx_*\succ x^*_{C\cup A_N\setminus A}x_*$. However, note that $x^*_{C\cup A_N\setminus A}x_*$ and $x^*_{C\cup A_N}x_*$ are equal to each other on the complement of $\sS\times A$. Since the latter set is null, we obtain $x^*_Bx_*\succ x^*_{C\cup A_N}x_*$ and, finally, $B\succ C\cup A_N$.
 
 \noindent  \textit{Step 5.}  \textit{$\succsim$ is atomless.}  Fix $A\in\sB_\sT$ such that $A\succ\varnothing$. Let $I^+=\{t\in \sT | A\succ A\cap [0,t) \}$ and $I^-=\{t\in \sT | A\cap [0,t) \succ \varnothing \}$. First, note that $I^+$ and $I^-$ are nontempty. Indeed, $\{A\cap [0,1/n),$ $n\geq 1\}$ is almost-vanishing. By Step 4, we obtain $A\succ A\cap [0,1/N)$  for some $N\geq 1$. Then, $1/N\in I^+$. Likewise, $\{[n,+\infty)$, $n\geq 1\}$ is vanishing. By \textsc{Monotone Continuity}, we obtain   $A\cap [0,M)\succ \varnothing$ for some $M\geq 1$. Then, $M\in I^-$.   Second, $I^+$ and $I^-$ are open in $\sT$. Indeed, fix $t\in I^+$. Then, $A\succ A\cap [0,t)$. Since $\{A\cap [t,t+1/n),$ $n\geq 1\}$ is almost-vanishing, there exists $N\geq 1$ such that $A\succ A\cap [0,t+1/N)$ by Step 4. Then, $t+1/N\in I^+$ and, by \textsc{$\sT$-Monotonicity}, $(t-1/N,t+1/N)\subseteq I^+$. Likewise, fix $t\in I^-$ so that  $A\cap [0,t) \succ \varnothing$. Since $\{[t-1/n,t)$, $n\geq 1\}$ is vanishing, \textsc{Monotone Continuity} yields the existence of $M\geq 1$ such that $A\cap [0,t-1/M) \succ \varnothing$. Then, $t-1/M\in I^-$ and, by \textsc{$\sT$-Monotonicity}, $(t-1/M,t+1/M)\subseteq I^-$. Finally, suppose first that $I^+$ and $I^-$ overlap. Then, let $t\in I^+\cap I^-$. We have  $A\succ A\cap [0,t) \succ \varnothing$ so that $A$ cannot be an atom. If $I^+$ and $I^-$ are disjoint, then, since $\sT$ is connected, there must exist $t\in\sT$ such that $t\notin I^+$ and $t\notin I^-$. By \textsc{$\sT$-Monotonicity} and (i), this implies $A\cap [0,t)\sim A$ and  $A\cap [0,t)\sim  \varnothing$, a contradiction.
\ethmprf

\Proposition{\label{prop:discounting}}
{There exists a unique countably additive and nonatomic probability measure $\eps_\lambda$ on $\sB_\sT$ with $\eps_\lambda[0,t)=1-e^{-\lambda t}=F_\lam(t)$ such that $A\succsim B$ if and only if $\eps_\lambda(A) \geq \eps_\lambda(B)$ for all $A,B\in\sB_\sT$.}

\bthmprf By Lemma \ref{lem:Qualitative-Prob}, $\succsim$ is a monotonely continuous and atomless qualitative probability on $\sB_\sT$ in the sense of \cite{VILL/64}. By his Theorem 3, Section 4, there exists a unique countably additive and nonatomic probability measure $\eps$ on $\sB_\sT$ providing a representation of $\succsim$. By Step 1 in the proof of Lemma \ref{lem:Qualitative-Prob}(v), for all $A,B\in\sB_\sT$ and $t\in\sT$, $A\succsim B$ if and only if $t+A\succsim t+B$. By  the uniqueness of the representation of $\succsim$ on $\sB_\sT$, we obtain: For all $A\in\sB_\sT$ and $t\in\sT$,
\beqn\label{cauchy}\eps(A)\ =\ \frac{\eps(t+A)}{\eps[t,+\infty)}.\eeqn
In particular, for $A=[t',+\infty)$, we obtain $\eps[t+t',+\infty)=\eps[t,+\infty)\cdot\eps[t',+\infty)$ for all $t,t'\in\sT$. By standard arguments, we must have $\eps[t,+\infty)=e^{-\lambda t}$ for all $t\in\sT$ and some $\lambda>0$. Then, by countable additivity and uniqueness in the Caratheodory extension theorem, we have $\eps=\eps_\lambda$.
\ethmprf

\noindent We conclude  Appendix \ref{AppA} showing how to construct time equivalents for acts $f\in \sF$ bounded by two outcomes.

\Lemma{\label{lem:time-equivalents}}
{For all $f\in\sF$ and $x,y\in\sX$ such that $x\succ f\succ y$, there exists $A\in\sB_\sT$ such that $f\sim {x}_Ay$. Moreover, we may assume $A=[0,t)$ for some $t\in\sT$.
}

\bthmprf  Remark that for all  $A,B\in\sB_\sT$ and $x,y\in\sX$ such that $x\succ y$, $A\succsim B$ if and only if $x_Ay\succsim x_By$. This follows noting that \textsc{$\sT$-Separability} implies the following form of Savage's (\citeyear{SAVA/54}) P4: For all $A,B\in\sB_\sT$ and $x,y,x',y'\in\sX$ such that $x\succ y$ and  $x'\succ y'$, $x_Ay\succsim x_By$ if and only if $x'_Ay'\succsim x'_By'$. See Machina and Schmeidler's (\citeyear{MACH/SCHM/92}) Section 4.2.

\noindent Now, let $I^-=\{\eps_\lambda(A)|f\succ {x}_A y, \, A\in\sB_\sT\}$. Clearly, $0\in I^-$ and $1\notin I^-$. Moreover, by \textsc{$\sT$-Monotonicity} and the nonatomicity of $\eps_\lambda$, if $q\in I^-$ and $q'\leq q$, then $q'\in I^-$. Now, fix $q\in I^-$. We will construct some $q'\in I^-$ such that $q'>q$. Since $q\in I^-$, there exists  $A\in\sB_\sT$ such that $q=\eps_\lambda(A)$ and $f\succ {x}_A y$. 
    Moreover, we must have $q<1$. Then, there exists $t\in\sT$ such that $F_\lam(t)=\eps_\lambda(A)$. Then, set $B=[0,t)\in\sB_\sT$. We have $\eps_\lambda(B)=\eps_\lambda(A)$. By Proposition \ref{prop:discounting} and the first part of the proof,    we obtain ${x}_A y\sim {x}_B y$ and hence $f\succ {x}_B y$. By \textsc{Monotone Continuity}, there exists $N\geq 1$ such that $f\succ {x}_{[N,+\infty)}({x}_By)={x}_{[N,+\infty)\cup B} y$. Therefore, we have $q'\in I^-$ where    $q'=\eps_\lam([N,+\infty)\cup B)$. Now, if $[N,+\infty)$ and $B$ are disjoint we have $q'>\eps_\lambda(B)=q$. If the two sets are not disjoint, then $[N,+\infty)\cup B=\sT$ and $1=q'\in I^-$, a contradiction. This shows that $I^-=[0,\underline{q})$ for some $\underline{q}\in (0,1)$.

\noindent Let $I^+=\{\eps_\lambda(A)|{x}_A y\succ f, \, A\in\sB_\sT\}$. Proceeding as in the previous paragraph, we obtain $\overline{q}\in (0,1)$ such that $I^+=(\overline{q},1]$.

\noindent Now, we must have $\underline{q}\leq\overline{q}$. Otherwise, consider any $q\in [0,1]$ such that $\underline{q}>q>\overline{q}$. Then, there exist $A,B\in\sB_\sT$ such that $\eps_\lambda(A)=\eps_\lambda(B)=q$ with  $f\succ {x}_A y$ and $ {x}_B y\succ f$. However, by Proposition \ref{prop:discounting} and the first part of the proof, $\eps_\lambda(A)=\eps_\lam(B)$ implies  ${x}_A y\sim {x}_B y$, a contradiction. Finally, take any  $q\in [0,1]$ such that $\underline{q}\leq q\leq\overline{q}$. Then, $q\notin I^-$ and $q\notin I^+$. 
By nonatomicity, there exists $A\in\sB_\sT$ such that $\eps_\lambda(A)=q$. Then,   $f\succsim {x}_A y$ and ${x}_A y\succsim f$.
   This implies $f\sim x_Ay$.
\ethmprf


\section{Utility and probability}\label{AppB}
\setcounter{thm}{0} \setcounter{equation}{0}
\renewcommand{\theequation}{B\arabic{equation}}
\renewcommand{\thethm}{B\arabic{thm}}

\medskip\noindent In Appendix \ref{AppB} we construct an \cite{ANSC/AUMA/63} setup. This allows us to derive a countably additive probability measure $\mu$ on $\sB_\sS$ and a nonconstant function $u$ from $\sX$ to $\Real$ using \cite{ANSC/AUMA/63}. Moreover, we prove our representation result for finitely valued acts.

\medskip\noindent We start by  introducing some notation. 

\begin{itemize}
    \item $\sX_0^\sT$ denotes the set of all finitely-valued and measurable deterministic acts. \item   $\sL$ denotes the set of all (finitely-supported) lotteries on $\sX$. 
    \item  $\lambda>0$ is as in Proposition \ref{prop:discounting}.
    \item $\phi:\sX_0^\sT\rightarrow \sL$ is defined 
    for all $\bx\in\sX_0^\sT$ and $x\in\sX$ by $\phi(\bx)(x)\ =\ \eps_\lambda\{\bx(\cdot)\ =\ x\}$. Note that, by the nonatomicity of $\eps_\lambda$, this mapping is surjective.
    \item $\sA=\{\alpha : \sS\rightarrow\sL | \alpha \text{ is finitely-valued and measurable} \}$. Measurability of $\alpha\in\sA$ means that $\alpha^{-1}(\{l\})\in\sB_\sS$ for all $l\in\sL$.
    \item  $\sF_0$ denotes the set of all acts $f\in\sF$ such that there exist a finite measurable partition $\Pi_\sS$ of $\sS$ and  a finite measurable partition $\Pi_\sT$ of $\sT$ such that $f$ is constant on $E_\sS\times E_\sT$ for all $E_\sS\in\Pi_\sS$ and  $E_\sT\in\Pi_\sT$. Note that $\sX_0^\sT \subset \sF_0$.
    \item For all $f\in\sF_0$,  $\varphi(f):\sS \rightarrow \sL$ is defined by  $\varphi(f)(s)(x)=\eps_\lambda\{ f(s,\cdot)=x\}$ for all $s\in\sS$ and $x\in\sX$. In words, $\varphi(f)(s)$ is the probability distribution induced by $f(s,\cdot)$ under $\eps_\lambda$.
    \item $\varphi: \sF_0\rightarrow \sA$ is surjective function from $\sF_0$ to $\sA$.
    \item Finally we remark  that for all $\bx\in\sX_0^\sT$, $f\in\sF_0$ and $s\in\sS$
  \beqn\label{varphiphi}\varphi(\bx)(s)=\phi(\bx)\quad\text{and}\quad \varphi(f)(s)=\phi(f(s,\cdot)).\eeqn
\end{itemize}

\medskip\noindent Note that $\sA$ is the standard \cite{ANSC/AUMA/63} framework.  For $l,m\in\sL$ and $\mu\in [0,1]$, we define the mixture $\mu l+(1-\mu)m\in\sL$ in the usual way by setting, for all $x\in\sX$, 
\[(\mu l+(1-\mu)m)(x)\\ =\ \mu l(x)+(1-\mu)m(x).\]
This mixture operation extends readily to $\sA$. For $\alp,\bet\in\sA$ and $\mu\in [0,1]$, let $\mu \alp+(1-\mu)\bet\in\sA$ by defined by, for all $s\in\sS$,
\[(\mu \alp+(1-\mu)\bet)(s)\\ =\ \mu \alp(s)+(1-\mu)\bet(s).\]

\medskip\noindent We will construct now a preference relation $\succsim_\sA$ on $\sA$ satisfying \cite{ANSC/AUMA/63} axioms. A binary relation $\succsim_\sA$ on $\sA$ is {\dfn monotonic} if, for all  $\alpha,\bet\in\sA$ such that $\alp(s)\succsim_\sA\bet(s)$ for all $s\in\sS$, we have $\alp\succsim_\sA\bet$.  We say that $\succsim_\sA$ satisfies {\dfn Independence} if, for all  $\alpha,\bet,\gam\in\sA$ and $\mu\in (0,1)$, $\alp\succsim_\sA \bet$ holds if and only if $\mu \alp+(1-\mu)\gam\succsim_\sA \mu\bet +(1-\mu)\gam$ holds. We say that $\succsim_\sA$ satisfies {\dfn Continuity} if, for all $\alp\in\sA$ and $x,y\in\sX$ such that $x\succsim_\sA \alp\succsim_\sA y$, there exists $\mu\in [0,1]$ such that $\alp\sim_\sA \mu x+(1-\mu)y$.

\Lemma{\label{lem:AA_mon}}
{There exists a nontrivial, complete, transitive and monotonic binary relation $\succsim_\sA$ on $\sA$ such that, for all $f,g\in\sF_0$,
$f \succsim g$ if and only if $\varphi(f) \succsim_\sA  \varphi(g)$. }

\bthmprf  \textit{Step 1.}  \textit{There exists a  nontrivial, complete and transitive binary relation $\succsim_\sL$ on $\sL$ such that, for all $\bx,\by\in\sX_0^\sT$, $\bx  \succsim_\sT \by$ if and only if $\phi(\bx) \succsim_\sL \phi(\by)$.}   The proof is similar to that of Machina and Schmeidler's (\citeyear{MACH/SCHM/92}) Theorem 1. However, note that they assume P6 while we do not. In fact, they only use P6 in their Step 1 to construct the probability measure.  We have already constructed the measure in Proposition \ref{prop:discounting} and do not need P6. The rest is identical.

\medskip\noindent \textit{Step 2.} \textit{For all $f,g\in\sF_0$, $\varphi(f)=\varphi(g)$ implies $f\sim g$.} Take $f,g\in\sF_0$ such that $\varphi(f)=\varphi(g)$.  By the second equality in Formula (\ref{varphiphi}), we have $\phi(f(s,\cdot))=\phi(g(s,\cdot))$ for all $s\in\sS$. By Step 1, we obtain $f(s,\cdot)\sim_\sT g(s,\cdot)$ for all $s\in\sS$. \textsc{Dominance} finally yields $f\sim g$.

\medskip\noindent We define $\succsim_\sA$ as follows: For all $\alpha,\beta\in\sA$, we set $\alpha\succsim_\sA\beta$ if and only if $f\succsim g$ for some $f,g\in\sF_0$ such that $\varphi(f)=\alpha$ and $\varphi(g)=\beta$. Preference $\succsim_\sA$ is well defined by Step 2. 

\medskip\noindent \textit{Step 3.}   \textit{$\succsim_\sA$ is nontrivial, complete, transitive and monotonic.} Nontriviality, completeness and transitivity of $\succsim_\sA$ follow from standard arguments. As for monotonicity, suppose $\alpha,\beta\in\sA$ are such that $\alpha(s)\succsim_\sA\beta(s)$ for all $s\in\sS$. The first equality in Formula (\ref{varphiphi}) implies  that $\succsim_\sL$ and $\succsim_\sA$ agree on $\sL$. So we have $\alpha(s)\succsim_\sL\beta(s)$ for all $s\in\sS$. The second equality in Formula (\ref{varphiphi})  and Step~1 then imply  $f(s,\cdot)\succsim_\sT g(s,\cdot)$ for all $s\in\sS$ where $f,g\in\sF_0$ are such that $\varphi(f)=\alpha$ and $\varphi(g)=\beta$. By \textsc{Dominance}, we obtain $f\succsim g$ and finally $\alpha\succsim_\sA\beta$.
\ethmprf

\Lemma{\label{lem:AA_ind_cont}}
{$\succsim_\sA$ satisfies Independence and Continuity.
}

\bthmprf \textit{Step 1.} \textit{For all $t\in\sT$ and $n\in\sL$, there exists $\bz\in\sX_0^\sT$ such that, for all $x\in\sX$, $\eps_\lambda[\{\bz(\cdot)=x\}\cap [0,t)]]= n(x)\cdot\eps_\lambda[0,t)$.} Let $\{x_1,\ldots, x_N\}\subseteq\sX$ be the support of $n$ and set $p_i=n(x_i)$ for all $i\in [1\ldots N]$. By the continuity of $F_\lam$, we can partition $[0,t)$ into intervals $[t_i,t_{i+1})$ for $i\in [0\ldots N]$ with $t_0=0$ and $t_{N}=t$ and also $F_\lam(t_{i+1})-F_\lam(t_i)=p_i\cdot \eps_\lambda[0,t)$ for all $i\in [1\ldots N]$. Then, it is sufficient to take any $\bz\in\sX_0^\sT$ constantly equal to $x_i$ on each $[t_i,t_{i+1})$.

\medskip\noindent  \textit{Step 2.} \textit{For all $t\in\sT$ and $\gamma\in\sA$, there exists $h\in\sF_0$ such that, for all $s\in\sS$ and $x\in\sX$, $\eps_\lambda[\{h(s,\cdot)=x\}\cap [0,t)]] = \gamma(s)(x)\cdot\eps_\lambda[0,t)$.}   Let $\Pi_\sS=\{E_i$, $i\in [1\ldots n]\}$ be a measurable partition of $\sS$ such that $\gamma$ is constantly equal to some $l_i\in\sL$ for all $ i\in [1\ldots n]$. By Step~1, we can find $\bz_i\in\sX_0^\sT$ such that, for all $x\in\sX$, 
\[\eps_\lam[\{\bz_i(\cdot)=x\}\cap [0,t)]]\ =\ l_i(x)\cdot\eps_\lam[0,t),\]for all $ i\in [1\ldots n]$. Now, each $\bz_i$ is finitely-valued. So there exists a finite measurable partition $\Pi_\sT$ of $\sT$ that is adapted to every $\bz_i$. We define a function $h$ from $\sS\times\sT$ to $\sX$ by setting $h(s,t)=\bz_i(t)$ where $i\in [1\ldots n]$ is such that $s\in E_i$. Clearly, $h$ is finitely-valued and adapted to $\Pi_\sS\times\Pi_\sT$. Hence, $h$ lies in $\sF_0$ and has the desired property.

\medskip\noindent \textit{Step 3.} \textit{$\succsim_\sA$ satisfies Independence}. Fix $\mu\in (0,1)$ and $\alp,\beta,\gamma\in\sA$. By the continuity of $F_\lam$, we can find $t\in\sT$ such that $1-\mu=\eps_\lam[0,t)$. Let $h\in\sF_0$ be as in the Lemma Step 2. Since $\varphi$ is surjective, we can find $f,g\in\sF_0$ such that $\varphi(f)=\alp$ and $\varphi(g)=\bet$. Next, for all $s\in\sS$ and $x\in\sX$, 
\beq
\varphi(h_tf)(s)(x)&=& \eps_\lambda[\{h_{t}f(s,\cdot)=x\}]\\
&=&(1-\mu)\cdot \eps_\lambda[\{h(s,\cdot)=x\}|[0,t)]]\ +\ \mu\cdot \eps_\lambda[t+\{f(s,\cdot)=x\}|[t,+\infty)]\\
&=& (1-\mu)\cdot   \gam(s)(x)\ +\ \mu \cdot \eps_\lambda[\{f(s,\cdot)=x\}]\\
&=& (1-\mu)\cdot  \gam(s)(x)\ +\ \mu \cdot\alp(s)(x),
\eeq where the third equality is by Step~2 and Formula (\ref{cauchy}). We obtain $\varphi(h_{t}f)=(1-\mu)\gam+\mu\alpha$. A similar argument provides $\varphi(h_{t}g)=(1-\mu)\gam+\mu\bet$.  Finally, we have 
\[\alp\ \succsim_\sA\ \bet\quad\Longleftrightarrow\quad f\ \succsim\ g\quad\Longleftrightarrow\quad h_tf\ \succsim\ h_tg,\]
where the first equivalence is by the definition of $\succsim_\sA$ in Lemma \ref{lem:AA_mon}, and the second one is by \textsc{Stationarity}. Then, Lemma \ref{lem:AA_mon} and the previous paragraph provide
\[\alp\ \succsim_\sA\ \bet\quad\Longleftrightarrow\quad \varphi(h_tf)\ \succsim_\sA\ \varphi(h_tg)\quad\Longleftrightarrow\quad (1-\mu)\gam+\mu\alp\ \succsim_\sA\ (1-\mu)\gam+\mu\bet.\]

\noindent \textit{Step 4.} \textit{$\succsim_\sA$ satisfies Continuity}. Suppose $\alp\in\sA$ and $x,y\in\sX$ are such that $x\succsim_\sA \alp\succsim_\sA y$.  If $\alp\sim_\sA x$ or $\alp\sim_\sA y$, we are done. So we may suppose $x\succ_\sA \alp\succ_\sA y$. Let $f\in\sF_0$ be such that $\varphi(f)=\alp$. Then, $x\succ f\succ y$. By Lemma \ref{lem:time-equivalents}, there exists $A\in\sB_\sT$ such that $f\sim x_Ay$. Then, $\varphi(x_Ay)=\alpha x+(1-\alpha)y$ with $\alpha=\eps_\lambda(A)$. Hence, we have $\alp\sim_\sA\alpha x+(1-\alpha)y$.
\ethmprf

\Proposition{\label{prop:rep_finite_act}}
{There exists a countably additive probability measure $\mu$ on $\sB_\sS$ and a nonconstant, measurable and bounded function $u$ from $\sX$ to $\Real$ such  that, for all $f,g\in\sF_0$,  \[f\ \succsim \ g\quad \Longleftrightarrow\quad \int_{\sS\times\sT} u[f(s,t)]d(\mu\times \eps_\lambda)(s,t)\quad\geq\quad\int_{\sS\times\sT} u[g(s,t)]d(\mu\times \eps_\lambda)(s,t).\]
Moreover, $\mu$ is unique and $u$ is unique up to positive affine transformation.
}

\bthmprf \textit{Step 1.} \textit{Representation of $\succsim$}. By Lemmata \ref{lem:AA_mon} and \ref{lem:AA_ind_cont}, we can apply the \cite{SCHM/89} version of the \cite{ANSC/AUMA/63} theorem and obtain a nonconstant mixture-linear function $v$ from $\sL$ to $\Real$ and a finitely-additive probability measure $\mu$ on $\sB_\sS$ such that, for all $\alpha,\beta\in\sA$,
\[\alp \ \succsim_\sA\ \bet\quad\Longleftrightarrow\quad E_\mu[v\circ\alp]\ \geq\ E_\mu[v\circ\bet].\]
 \textsc{Monotone Continuity} implies  the countable additivity of $\mu$. For instance, see \cite{ARRO/70}. Moreover, let $u$ denote the function from $\sX$ to $\Real$ obtained as the restriction of $v$ to $\sX$. By mixture-linearity, we have $v(l)\ =\ E_l[u]$  for all $l\in\sL$. Then, for all $f,g\in\sF_0$, we have
\[f \ \succsim\ g\quad\Longleftrightarrow\quad E_\mu[E_{\varphi(f)(\cdot)}[u]]\ \geq\ E_\mu[E_{\varphi(g)(\cdot)}[u]].\]
From there, the representation of $\succsim$ on $\sF_0$ follows from the remark that, for all $f\in\sF_0$ 
\[E_\mu[E_{\varphi(f)(\cdot)}[u]]\ =\ \int_{\sS}\left(\int_\sT u[f(s,t)]d\eps_\lambda(t)\right)d\mu(s) \ =\ \int_{\sS\times\sT} u[f(s,t)]d(\mu\times \eps_\lambda)(s,t), \]
where the second equality invokes the Fubini theorem.

\noindent \textit{Step 2.} \textit{$u$ is measurable.}  For every $q\in\mathbb{R}$, we show that $U:=u^{-1}(q,+\infty)\in\sB_\sX$. If $u(x)> q$ for all $x\in\sX$, then $U=\sX\in\sB_\sX$. If $u(x)\leq q$ for all $x\in\sX$, then $U=\varnothing\in\sB_\sX$. In the remaining case, we have $u(x)\leq q <u(y)$ for some $x,y\in\sX$. We can then find $\alpha\in [0,1)$ such that $q=\alpha u(y)+(1-\alpha)u(x)$. Let $t\in\sT$ be such that $F_\lambda(t)=\alpha$, and set $\bx=y_tx\in\sX^\sT_0$. We have $\int_\sT u[\bx(t)]d\eps_\lambda(t)=q$ and, by Step 1,  $U=\{z\in\sX$, $z\succ \bx\}$. Then, $U\in\sB_\sX$ by \textsc{$\sT$-Measurability}. The measurability of $u^{-1}(-\infty,q)$ can be proved likewise.

\noindent \textit{Step 3.} \textit{$u$ is bounded.} Suppose by way of contradiction that $u$ is unbounded from above. Consider a partition $\{A_n$, $n\geq 1\}$ of $\sT$ such that the $A_n$ are successive intervals with $A_1$ starting from $t=0$ and such that $\eps_\lambda(A_n)=1/2^n$ for all $n\geq 1$. Let $\bx\in \sX^\sT$ be such that the value of $\bx$ over $A_n$ is an outcome $x_n\in \sX$ such that $u(x_n)\geq 2^n$. Such an outcome exists since $u$ is unbounded from above.  We may suppose that $u(x_{n+1})>u(x_n)$ for all $n\geq 1$ without loss of generality.

\noindent Note that we have for all $N\geq 1$
\[\sum_{n=1}^{+\infty}\text{min}(u(x_n),u(x_N))\cdot\eps(A_n)\ \geq\ \sum_{n=1}^N u(x_n)\cdot\eps(A_n)\ \geq\ N\]

\noindent We prove that  $\bx\succsim z$ for all $z\in \sX$. Fix $z\in \sX$. By the previous formula, there exists $N\geq 1$  such that 
\[\sum_{n=1}^{+\infty}\text{min}(u(x_n),u(x_N))\cdot\eps(A_n)\ \geq\ u(z).\] Then, set $y=x_N$ and let $\by\in\sX^\sT$ be such that, for all $t\in\sT$, $\by(t)=\bx(t)$ if $y\succsim \bx(t)$ and  $\by(t)=y$ if $\bx(t)\succ y$. Since $u$ and $\bx$ are measurable, so is $\by$. Furthermore, by \textsc{$\sT$-Monotonicity}, $\bx\succsim\by$ Furthermore, $\by\in\sX^\sT_0$. The previous formula and the representation on $\sX^\sT_0$ obtained in Step 1 yield $\by\succsim z$ and, finally, $\bx\succsim z$.

\noindent Define now $\bz\in\sX^\sT$ by $\bz(t)=\bx(t)$ for all $t\in\sT\setminus A_1$ and $\bz(t)=x_2$ for all $t\in A_1$. Since $u$ and $\bx$ are measurable, so is $\bz$.  We have $\eps_\lambda(A_1)>0$ and $A_1$ is hence non-null by Step 1. Then, \textsc{$\sT$-Monotonicity} yields $\bz\succ \bx$. Since $\{\cup_{i\geq n}A_i$, $n\geq 1\}$ is a vanishing sequence, by \textsc{Monotone Continuity} there is $n\geq 2$ such that $\bz_{t_n}x_1\succ \bx$, where $t_n\in\sT$ denotes the lower bound of $A_n$.
Note that $\bz_{t_n}x_1$ lies in $\sX^\sT_0$. The preferred outcome in its range is $x_{n-1}$. By \textsc{$\sT$-Monotonicity}, $x_{n-1}\succsim \bz_{t_n}x_1$ and, therefore, $x_{n-1}\succ\bx$, which contradicts the previous paragraph. Therefore $u$ must be bounded from above. A similar argument shows that it is also bounded from below.
\ethmprf


\section{Representation}\label{AppC}
\setcounter{thm}{0} \setcounter{equation}{0}
\renewcommand{\theequation}{C\arabic{equation}}
\renewcommand{\thethm}{C\arabic{thm}}

In Appendix \ref{AppC} we prove sufficiency of the axioms for the representation in Theorem \ref{main}. First, we prove the representation for bounded deterministic acts in $\sX^\sT$ (Lemma \ref{lem:repr_time_acts_bounded}), then for bounded acts in $\sF$ (Lemma \ref{lem:repr_acts_bounded}), and finally for general acts in $\sF$ (Proposition \ref{prop:rep_all_act}).

\medskip\noindent We say that   $f\in\sF$ is {\dfn bounded} if there exist $x_0,x_1\in\sX$ such that $x_1 \succsim f(s,t) \succsim x_0$ for all $s\in\sS$ and $t\in\sT$. Proposition \ref{prop:rep_finite_act} shows that $u$ is bounded and measurable. We may then define functions $V:\sF\rightarrow\Real$ and $U:\sX^\sT\rightarrow\Real$ by
\[V(f)\ =\ \int_{\sS\times\sT}u[f(s,t)]d(\mu\times\eps_\lam)(s,t)\quad\text{and}\quad U(\bx)\ =\ \int_{\sT}u[\bx(t)]d\eps_\lam(t).\]

\Lemma{\label{lem:time_equivalents2}}
{For all bounded $f\in\sF$,  there exists $\bx\in\sX^\sT_0$ such that $f\sim\bx$.
}
\bthmprf Suppose that $f\in\sF$ is such that $x \succsim f(s,t) \succsim y$ for all $s\in\sS$ and $t\in\sT$ and for some $x,y\in\sX$. Then, by \textsc{$\sT$-Monotonicity}, $x \succsim_\sT f(s,\cdot) \succsim_\sT y$ for all $s\in\sS$. By \textsc{Dominance}, $x \succsim f \succsim y$. If $f\sim x$ or $f\sim y$, we are done. So we may suppose $x \succ f \succ y$. Then, the result follows from Lemma \ref{lem:time-equivalents}.
\ethmprf

\medskip\noindent For $A,B\in\sB_\sT$ and $\bx\in\sX^\sT$, we write $A\perp B$ if $\eps_\lam(A\cap B)=\eps_\lam(A)\cdot\eps_\lam(B)$ and $\bx\perp A$ if $\{t\in\sT$, $\bx(t)=x\}\perp A$ for all $x\in\sX$.

\Lemma{\label{lem:P7-like}}
{For all $\bx\in\sX^\sT$, all finite measurable partition $\{A_1,\ldots, A_N\}$ of $\sT$ and sequence $\{\bx_1,\ldots,\bx_N\}$ of elements of $\sX_0^\sT$ such that $\bx_n\perp A_m$ for all $n,m\in [1\ldots N]$,

\bdesc
\item[(i)] If $\bx(t)\succsim \bx_{n}$ for all $t\in A_n$ and all $n\in [1\ldots N]$, then $\bx  \succsim \sum_{n=1}^N\bone_{A_n}\bx_{n}$,
\item[(ii)] If $\bx_{n}\succsim\bx(t)$ for all $t\in A_n$ and all $n\in [1\ldots N]$, then $\sum_{n=1}^N\bone_{A_n}\bx_{n}\succsim\bx$.
\edesc  

}

\bthmprf We only show (i). Consider first the case where $\bx$ is constant on each cell of the partition, i.e. $\bx=\sum_{n=1}^N\bone_{A_n}x_n$ with $x_n\in\sX$ for all $n\in [1\ldots N]$. Then, by assumption, we have $x_n\succsim \bx_{n}$ and Proposition \ref{prop:rep_finite_act} implies $u(x_n)\geq U(\bx_{n})$ for all $n\in [1\ldots N]$. We obtain 
\[U(\bx)\ =\ \sum_{n=1}^N\eps_\lambda(A_n)\cdot u(x_n)\ \geq  \sum_{n=1}^N\eps_\lambda(A_n)\cdot U(\bx_{n})\ =\ U(\sum_{n=1}^N\bone_{A_n}\bx_{n}),\]
where the last equality is because $\bx_{n}\perp A_n$ for all $n\in [1\ldots N]$. Since $\bx$ is an element of $\sX_0^\sT$, Proposition \ref{prop:rep_finite_act} yields the desired ranking.

\noindent Consider next the case where $\bx$ has a minimum on each $A_n$, i.e. for each $n\in [1\ldots N]$, there exists $t_n\in A_n$ such that $\bx(t)\succsim \bx(t_n)$ for all $t\in A_n$ and set $x_n=\bx(t_n)$. Then, by \textsc{$\sT$-Monotonicity}, we have $\bx\succsim \sum_{n=1}^N\bone_{A_n}x_{n}$. Since $x_n\succsim \bx_{n}$ for all $n\in [1\ldots N]$, we can apply the previous paragraph and obtain $\sum_{n=1}^N\bone_{A_n}x_{n}\succsim \sum_{n=1}^N\bone_{A_n}\bx_{n}$. Transitivity allows to conclude.

\noindent Consider finally the general case and suppose by contradiction that $\sum_{n=1}^N\bone_{A_n}\bx_{n}\succ\bx$. Let $I$ collect all integers $n\in [1\ldots N]$ such that $\bx$ has no minimum on $A_n$. Fix any $m\in I$.  Let $\{t_n^m$, $n\geq 1\}$ be a sequence of points in $\sT$ such that $\{u[\bx(t_n^m)]$, $n\geq 1\}$ is decreasing and converges to $\text{inf}_{A_m}u\circ\bx$. For all $n\geq 1$, define $B_n^m=\{t\in A_m | u[\bx(t)]<u[\bx(t_n^m)]\}$. Since $\bx$ has no minimum on $A_m$, the sequence $\{B_n^m$, $n\geq 1\}$ is vanishing. Then, the sequence $\{C_n$, $n\geq 1\}$ defined by $C_n=\cup_{m\in I}B_n^m$ for all $n\geq 1$ is also vanishing. \textsc{Monotone Continuity} then yields $ \sum_{n=1}^N\bone_{A_n}\bx_{n}\succ x_{C_N}\bx$ for some $N\geq 1$ where $x\in\sX$ is a preferred outcome in the collection $\{\bx(t_1^m)$, $m\in I\}$. Then, set $\bx'=x_{C_N}\bx$. Observe first that, by construction, we have $\bx'(t)\succsim\bx(t)$ for all $t\in\sT$ and therefore obtain $\bx'(t)\succsim \bx_n$ for all $t\in A_n$ and $n\in [1\ldots N]$.
Indeed, if $t\in C_N$, then $t\in B_N^m$ for some $m\in I$ so that $u(\bx(t))\leq u[\bx(t_N^m)]\leq u(x)=u(\bx'(t))$ and therefore $\bx'(t)\succsim\bx(t)$ by Proposition \ref{prop:rep_finite_act}. Observe also that $\bx'$ has a minimum on each cell $A_m$. Indeed, if $m\notin I$, then $\bx'=\bx$ on $A_m$ with $\bx$ presenting a minimum on $A_m$ by definition of $I$. If $m\in I$, consider any $t\in A_m$. If $t\notin C_N$, then it must be that $t\notin B_N^m$ and $u[\bx'(t)]=u[\bx(t)]\geq u[\bx(t_N^m)]$ which by Proposition \ref{prop:rep_finite_act} yields $\bx'(t)\succsim \bx(t_N^m)$. Moreover if  $t\in C_N$, $\bx'(t)=x\succsim \bx(t_1^m)\succsim\bx(t_N^m)$. 
Overall, we may apply the previous paragraph to $\bx'$ and obtain $\bx'  \succsim \sum_{n=1}^N\bone_{A_n}\bx_{n}$ in contradiction with the ranking implied by \textsc{Monotone Continuity} above.
\ethmprf

\Lemma{\label{lem:repr_time_acts_bounded}}
{For all bounded $\bx,\by\in\sX^\sT$, $\bx \succsim_\sT  \by$ if and only if  $U(\bx)\geq U(\by)$. 
}

\bthmprf Consider first $\bx\in\sX^\sT$ and $\bx_0\in\sX^\sT_0$ such that $\bx\sim_\sT \bx_0$ with $\bx$ bounded. Existence of such an $\bx_0$ is guaranteed by Lemma \ref{lem:time_equivalents2}. We will show
 \beqn\label{aux.1}\int_{\sT} u[\bx(t)]d\eps_\lambda(t)  \ =\ \int_\sT u[\bx_0(t)]d\eps_\lambda(t).\eeqn
 
\noindent Let $x_0,x_1\in\sX$ be such that $x_1 \succsim \bx(t) \succsim x_0$ for all $t\in\sT$.  By applying positive affine transformations if necessary, we may assume $u(x_1)=1$ and $u(x_0)=0$ without loss of generality. Fix also $N\geq 1$.

\medskip\noindent For all $t\in\sT$, we have $1\geq u[\bx(t)]\geq 0$. Let $\Pi_\sT=\{A_1,\ldots,A_N\}$ be the partition of $\sT$ defined for all $n\in [1\ldots N]$ by 
\[A_n\ =\ \left\{t\in\sT,\ \frac{n-1}{N}\ \leq \ u[\bx(t)]\  < \ \frac{n}{N}\right\}\]
By the measurability of  $u$ and $\bx$, $\Pi_\sT$ forms a finite measurable partition of $\sT$ possibly with empty subsets. Then, by the monotonicity of the integral, we further obtain 
\beqn\label{ext.eq1.timeis}\sum_{n=1}^N\eps_\lambda(A_n)\cdot \frac{n-1}{N}\quad \leq\quad    \int_{\sT} u[\bx(t)]\ \textit{d} \eps_\lambda(t) \quad\leq\quad \sum_{n=1}^N\eps_\lambda(A_n)\cdot \frac{n}{N}.\eeqn
Fix any $n\in [0\ldots N]$. Since $\eps_\lambda$ is countably additive and nonatomic, there exists $B_n\in\sB_\sT$ such that 
\[\frac{n}{N}\quad =\quad \eps_\lambda(B_n)\quad =\quad \int_{\sT}u[\bx_n]\ \textit{d}\eps_\lambda,\]
where $\bx_n={x_1}_{B_n}x_0\in\sX_0^\sT$. We can  assume $A\perp B_n$ for all $A\in\Pi_\sT$ without loss of generality. Indeed, for all $A\in\Pi_\sT$, the nonatomicity of $\eps_\lambda$ provides $B_n^A\in\sB_\sT$ such that $B_n^A\subseteq A$ and $\eps_\lambda(B_n^A)=(n/N)\cdot\eps_\lambda(A)$. Then, set $B_n=\cup_{A\in\Pi_\sT}B_n^A$. We have $B_n\in\sB_\sT$ and $\eps_\lam(B_n)=n/N$. Furthermore, for all $A\in\Pi_\sT$, we have $B_n\cap A=B_n^A$ and, therefore, $\eps_\lambda(B_n\cap A)=\eps_\lam(B_n)\cdot\eps_\lam(A)$.

 \noindent Next, we define $\by_0,\bz_0\in\sX_0^\sT$ in the following way:
\[\by_0\ =\ \sum_{n=1}^N\bone_{A_n}\bx_{n-1}\ \text{ and }\ \bz_0\ =\ \sum_{n=1}^N\bone_{A_n}\bx_{n}\]
Now, we have 
 \[\int_\sT u[\by_0(t)]d\eps_\lambda(t)\ =\ \sum_{n=1}^N\ \eps_\lambda(A_n\cap B_{n-1})\ =\ \sum_{n=1}^N\eps_\lambda(A_n)\cdot \frac{n-1}{N},\] and likewise for $\bz_0$. Furthermore, we have  by construction $\bx(t)\succsim \bx_{n-1} ={x_1}_{B_{n-1}}x_0$ for all $t\in A_n$ and $n\in [1\ldots N]$, and $\bx_{n-1}\perp A$ for all $n\in [1\ldots N]$ and $A\in\Pi_\sT$.  By Lemma \ref{lem:P7-like}, we obtain $\bx\succsim \by_0$. Similarly, we can prove that $\bz_0\succsim \bx$. Since $\bx\sim \bx_0$, we obtain  $\bz_0\succsim \bx_0\succsim\by_0$, 
which implies by Proposition \ref{prop:rep_finite_act}
 \beqn\label{ext.eq2.timeis}\sum_{n=1}^N\eps_\lambda(A_n)\cdot \frac{n-1}{N}\quad \leq\quad    \int_{\sT} u[\bx_0(t)]\ \textit{d} \eps_\lambda(t) \quad\leq\quad \sum_{n=1}^N\eps_\lambda(A_n)\cdot \frac{n}{N}.\eeqn
Combining Formulae (\ref{ext.eq1.timeis}) and  (\ref{ext.eq2.timeis}) gives
\[\big|\int_{\sT} u[\bx(t)]\textit{d}\eps_\lambda(t)\ -\ \int_{\sT} u[\bx_0(t)]\textit{d}\eps_\lambda\ \big|\ \leq \ \sum_{n=1}^N\eps_\lambda(A_n)\cdot \frac{n}{N}\ -\ \sum_{n=1}^N\eps_\lambda(A_n)\cdot \frac{n-1}{N}\ =\ \frac{1}{N}.\]
We finally obtain Formula (\ref{aux.1}) by taking the limit as $N$ goes to $\infty$. 

\noindent Consider next any bounded $\bx,\by\in\sX^\sT$. By Lemma \ref{lem:time_equivalents2}, there exist $\bx_0,\by_0\in\sX^\sT_0$ such that $\bx\sim_\sT \bx_0$ and $\by\sim_\sT \by_0$.  Finally, the result follows by Formula (\ref{aux.1}) and Proposition~\ref{prop:rep_finite_act}.
\ethmprf

\Lemma{\label{lem:repr_acts_bounded}}
{For all  bounded $f,g\in\sF$, $f\succsim g$ if and only if $V(f)\geq V(g)$. 
}

\bthmprf  Suppose  that $f\in\sF$ is bounded. By Lemma \ref{lem:time_equivalents2} there exists $\bx\in\sX^\sT_0$  such that $f\sim\bx$. We will show
\beqn\label{aux.2}\int_{\sS\times\sT} u[f(s,t)]d(\mu\times \eps_\lambda)(s,t)\quad =\quad \int_{\sT} u[\bx(t)]d\eps_\lambda(t).\eeqn
Let  $x_0,x_1\in\sX$ be such that $x_1 \succsim f(s,t) \succsim x_0$ for all $s\in\sS$ and $t\in\sT$. By applying positive affine transformations if necessary, we may assume $u(x_1)=1$ and $u(x_0)=0$ without loss of generality. Fix also $N\geq 1$.

\medskip\noindent For all $s\in\sS$ and $t\in\sT$, we have $1\geq u[f(s,t)]\geq 0$. Let $\Pi_\sS=\{E_1,\ldots,E_N\}$ be the partition of $\sS$ defined for all $n\in [1\ldots N]$ by 
\[E_n\ =\ \left\{s\in\sS,\ \frac{n-1}{N}\ \leq \ \int_{\sT}\ u[f(s,t)]\ \textit{d}\eps_\lambda(t)\ < \ \frac{n}{N}\right\}\]
Note that $\Pi_\sS$ forms a finite measurable partition of $\sS$ possibly with empty subsets. By the monotonicity of the integral, we further obtain 
\beqn\label{ext.eq1}\sum_{n=1}^N\mu(E_n)\cdot \frac{n-1}{N}\quad \leq\quad    \int_{\sS\times\sT} u[f(s,t)]\ \textit{d}(\mu\times \eps_\lambda)(s,t) \quad\leq\quad \sum_{n=1}^N\mu(E_n)\cdot \frac{n}{N}\eeqn
Now fix  $n\in [0\ldots N]$. By the nonatomicity of $\eps_\lambda$, there exists $A_n\in\sB_\sT$ such that 
\[\frac{n}{N}\quad =\quad \eps_\lambda(A_n)\quad =\quad \int_{\sT}u( \bx_n)\ \textit{d}\eps_\lambda,\]
where $\bx_n={x_1}_{A_n}x_0\in\sX_0^\sT$. Then, we define $f_0,g_0\in\sF_0$ in the following way:
\[f_0\ =\ \sum_{n=1}^N\bone_{E_n}\bx_{n-1}\ \text{ and }\ g_0\ =\ \sum_{n=1}^N\bone_{E_n}\bx_{n}\]
Now, fix $s\in\sS$ and let  $n\in [1\ldots N]$ be such that $s\in E_n$. Then, $f_0(s,\cdot)=\bx_{n-1}$ and $g_0(s,\cdot)=\bx_{n}$ and therefore
\[\int_{\sT} u[f_0(s,\cdot)]\textit{d}\eps_\lambda\ =\ \frac{n-1}{N}\ \leq \ \int_{\sT}\ u[ f(s,\cdot)]\ \textit{d}\eps_\lambda \ \leq \ \frac{n}{N} \ =\ \int_{\sT} u[ g_0(s,\cdot)]\textit{d}\eps_\lambda.\] 
Since $f(s,\cdot)\in\sX^\sT$ is bounded, Lemma \ref{lem:repr_time_acts_bounded}  yields $g_0(s,\cdot)\succsim_\sT f(s,\cdot)\succsim_\sT f_0(s,\cdot)$, and this holds for all $s\in\sS$. Then, \textsc{Dominance} further yields $g_0\succsim f\succsim f_0$. Since $f\sim \bx$, we obtain $g_0\succsim \bx\succsim f_0$. By Proposition \ref{prop:rep_finite_act} , we have:
\beqn\label{ext.eq2}\sum_{n=1}^N\mu(E_n)\cdot \frac{n-1}{N}\quad \leq\quad    \int_{\sT} u( \bx)\ \textit{d}\eps_\lambda \quad\leq\quad \sum_{n=1}^N\mu(E_n)\cdot \frac{n}{N}.\eeqn
Combining Formulae (\ref{ext.eq1}) and  (\ref{ext.eq2}) gives
\[\big|\int_{\sS\times\sT} u( f ) \textit{d}\mu\times \eps_\lambda\ -\ \int_{\sT}u( \bx) \textit{d}\eps_\lambda\ \big|\ \leq \ \sum_{n=1}^N\mu(E_n)\cdot \frac{n}{N}\ -\ \sum_{n=1}^N\mu(E_n)\cdot \frac{n-1}{N}\ =\ \frac{1}{N}.\]
We obtain Formula (\ref{aux.2}) taking the limit as $N$ goes to $\infty$.

\noindent Finally, consider  bounded $f,g\in\sF$. By Lemma \ref{lem:time_equivalents2}, there exist $\bx,\by\in\sX^\sT_0$ such that $f\sim_\sT \bx$ and $g\sim_\sT \by$.  The result follows by Formula (\ref{aux.2}) and Proposition \ref{prop:rep_finite_act}.
\ethmprf

 \medskip\noindent We say that $f\in\sF$ is {\dfn  bounded from above} if there exists $x\in\sX$ such that $x\succsim f(s,t)$ for all $(s,t)\in\sS\times\sT$. We say it is {\dfn bounded from below} if  there exists $y\in\sX$ such that $f(s,t)\succsim y$ for all $(s,t)\in\sS\times\sT$.

\Lemma{\label{lem:null_sets}}
{For all $E\in\sB$, $E$ is null if and only if $(\mu\times\eps_\lambda)(E)=0$.
}
\bthmprf Suppose $(\mu\times\eps_\lambda)(E)=0$. Let $f,g\in\sF$ be  such that $f=g$ on the complement of $E$. If $f$ and $g$ are bounded, then $V(f)=V(g)$, and $f\sim g$ follows from Lemma \ref{lem:repr_acts_bounded}.

\noindent Suppose now $f$ and $g$ are bounded from above, and not bounded from below. Let $\{x_n$, $n\geq 1\}$ be a sequence in $\sX$ such that $\{u(x_n)$, $n\geq 1\}$ is decreasing and converges to $\text{inf} \ u$. For all $n\geq 1$, let $E_n\in\sB$ be the collection of all $(s,t)\in\sS\times\sT$ such that $u[f(s,t)]<u(x_n)$ and $u[g(s,t)]<u(x_n)$. Since $f$ (or $g$) is not bounded from below, $\{E_n$, $n\geq 1\}$ is vanishing.  For all $n\geq 1$, let $f_n={x_n}_{E_n}f$ and $g_n={x_n}_{E_n}g$. Finally, suppose $f$ and $g$ are not indifferent to each other. Without loss of generality, we may suppose $f\succ g$. Then, by \textsc{Monotone Continuity}, we have $f\succ {x_1}_{E_N}g$ for some $N\geq 1$. By \textsc{$\sT$-Monotonicity} and \textsc{Dominance}, we obtain $f\succ {x_N}_{E_N}g$. Meanwhile, and still by \textsc{$\sT$-Monotonicity} and \textsc{Dominance}, we have ${x_N}_{E_N}f\succsim f$ and obtain $f_N\succ g_N$.  However, note that $f_N$ and $g_N$ are bounded and equal to each other except on $E_N^c\cap E$ with $(\mu\times\eps_\lambda)(E_N^c\cap E)=0$. Then, Lemma \ref{lem:repr_acts_bounded} gives $f_N\sim g_N$, hence a contradiction. 

\noindent Suppose next $f$ and $g$ are not bounded from above. Let $\{x_n$, $n\geq 1\}$ be a sequence in $\sX$ such that $\{u(x_n)$, $n\geq 1\}$ is increasing and converges to $\text{sup} \ u$. For all $n\geq 1$, let $E_n\in\sB$ be the collection of all $(s,t)\in\sS\times\sT$ such that $u[f(s,t)]>u(x_n)$ and $u[g(s,t)]>u(x_n)$. Supposing again $f\succ g$, we obtain  ${x_N}_{E_N}f\succ {x_N}_{E_N}g$ for some $N\geq 1$, which contradicts the two previous paragraphs since ${x_N}_{E_N}f$ and ${x_N}_{E_N}g$ are bounded from above and equal to each other on $E_N^c\cap E$ with $(\mu\times\eps_\lambda)(E_N^c\cap E)=0$.

\noindent Suppose finally that $E$ is null. Let $x,y\in\sX$ be such that $x\succ y$. Then,  $x_Ey$ and $y$ agree on the complement of $E$. Hence,  $x_Ey\sim y$. The two acts are bounded. By Lemma \ref{lem:repr_acts_bounded}, we obtain $(\mu\times\eps_\lambda)(E)=0$.
\ethmprf

\Lemma{\label{lem:double_mono}}
{For all $f,g\in\sF$, if $f(s,t)\succsim g(s,t)$ for all $s\in\sS$ and $t\in\sT$, then $f\succsim g$; if additionally,  $f(s,t)\succ g(s,t)$ for all $(s,t)$ in some non-null subset in $\sB$, then $f\succ g$.
}

\bthmprf Suppose first that $f,g\in\sF$ are such that $f(s,t)\succsim g(s,t)$ for all $s\in\sS$ and $t\in\sT$. Then, by \textsc{$\sT$-Monotonicity}, we have $f(s,\cdot)\succsim_\sT g(s,\cdot)$ for all $s\in\sS$. By \textsc{Dominance}, we further obtain $f\succsim g$. Suppose now additionally that  $f(s,t)\succ g(s,t)$ for all $(s,t)\in E$ with $E\in\sB$ non-null. By Lemma \ref{lem:null_sets}, we have $(\mu\times\eps_\lambda)(E)>0$. For all $s\in\sS$, let $E_s=\{t\in\sT$, $(s,t)\in E\}\in\sB_\sT$. Since we have
\[0<(\mu\times\eps_\lambda)(E)\ =\ \int_\sS\eps_\lambda(E_s)d\mu(s)\]
 it must be that the set $A =\{s\in\sS|\eps_\lambda(E_s)>0 \}\in\sB_\sS$  satisfies $\mu(A)>0$. Then, by Lemma~\ref{lem:null_sets}, $A$ is non-null. Furthermore, by definition of the set $A$ and by Lemma \ref{lem:null_sets}, $E_s$ is non-null for all $s\in A$. Now, fix $s\in A$. For all $t\in E_s$, we have $(s,t)\in E$ and therefore $f(s,t)\succ g(s,t)$. Then, since $E_s$ is non-null, \textsc{$\sT$-Monotonicity} implies $f(s,\cdot)\succ_\sT g(s,\cdot)$. The latter holds for all $s\in A$ with $A$ non-null. Then, \textsc{Dominance} yields $f\succ g$.
\ethmprf

\Proposition{\label{prop:rep_all_act}}
{For all  $f,g\in\sF$, $f\succsim g$ if and only if $V(f)\geq V(g)$. 
}
\bthmprf \textit{Step 1.} \textit{For all  $E\in\sB$ and $f\in\sF$, there exist $x,y\in\sX$ such that $x_Ef\succsim f\succsim y_Ef$.} Note that the result is straightforward if $E$ is null. Hence, we suppose that $E$ is non-null. \\
We only show the existence of $y\in\sX$ such that $ f\succsim y_Ef$. Suppose by contradiction that no such $y$ exists. Then $y_Ef\succ f$ for all $y\in\sX$.  If $f$ has a minimum on $E$, in the sense that  there exists $x\in\sX$ such that $f(s,t)\succsim x$ for all $(s,t)\in E$ with $x=f(s,t)$ for some $(s,t)\in E$, then, by Lemma \ref{lem:double_mono}, we have $f\succsim x_Ef$. However by our hypothesis we have $x_Ef\succ f$, a contradiction.\\
Suppose now that $f$ has no minimum on $E$ in the previous sense. Let $\{(s_n,t_n)$, $n\geq 1\}$ be a sequence in $E$ such that $\{u[f(s_n,t_n)]$, $n\geq 1\}$ is decreasing and converges to $\text{inf}\ u( f)$. For all $n\geq 1$, let $E_n\in\sB$ be the collection of all $(s,t)\in E$ such that $u[f(s,t)]\leq u[f(s_n,t_n)]$. Let also $F_n:=E\setminus E_n$ and $x_n:=f(s_n,t_n)$. Since $f$ has no minimum on $E$, $\{E_n$, $n\geq 1\}$ is vanishing. Then, for some $N\geq 1$, $F_N$ is non-null. Indeed, suppose that $F_n$ is null for all $n\geq 1$. Consider $f,g\in\sF$ such that $f=g$ on $\sS\setminus E$. If $f\succ g$,  \textsc{Monotone Continuity} yields $f\succ f_{E_N}g$ for some $N\geq 1$. Since $F_N$ is null, we further have $f\succ f_Eg$ with $f_Eg=f$, a contradiction. This shows that $f\sim g$ and hence that $E$ must be null, another contradiction. Then, we have $f(s,t)\succsim ({x_N}_{F_N} f)(s,t)$  for all $(s,t)\in\sS\times\sT$ with $f(s,t)\succ ({x_N}_{F_N} f)(s,t)$ for all $(s,t)\in F_N$ with $F_N$ non-null. By Lemma \ref{lem:double_mono}, we obtain $f\succ {x_N}_{F_N} f$. By \textsc{Monotone Continuity} there exists $M\geq N$ such that $f\succ f'$ where $f':={x_M}_{E_M}({x_N}_{F_N}f)$. Meanwhile, we also have $f'(s,t)\succsim x_M$  for all $(s,t)\in E$ with $f'=f$ on $\sS\setminus E$ and hence obtain $f'\succsim {x_M}_Ef$ by Lemma \ref{lem:double_mono}. However, we assumed  $y_Ef\succ f$ for all $y\in\sX$. Hence we  obtain $f'\succ f$, a contradiction.

\noindent \textit{Step 2.} \textit{For all $f\in\sF$, sequence $\{f_n$, $n\geq 1\}$ in $\sF$ and vanishing sequence $\{E_n$, $n\geq 1\}$ in $\sB$, if $f=f_n$ on $\sS\times\sT\setminus E_n$ for all $n\geq 1$, then $\{V(f_n)$, $n\geq 1\}$ converges to $V(f)$.}    For all $n\geq 1$ , we have
\[\big|V(f_n)\ -\ V(f)\big| \ =\ \big|\int_{\sS\times\sT}\bone_{E_{n}}\left(u( f_n)-u( f)\right)d(\mu\times\eps_\lambda)\big| \ \leq\ 2\cdot\text{sup} \ \big| u\big|\cdot (\mu\times\eps_\lam)(E_n).\]
Now, since $\{E_n$, $n\geq 1\}$ is vanishing and $\mu\times\eps_\lam$ is countably additive, the sequence $\{(\mu\times\eps_\lam)(E_n)$, $n\geq 1\}$ converges to $0$.

\noindent \textit{Step 3.} \textit{Representation of $\succsim$.} Fix $f\in\sF$. By Step 1, we can find $x,y\in\sX$ such that $x\succsim f\succsim y$. Then there exists $A\in\sB_\sT$ such that $f\sim x_Ay$. Indeed, this is obvious if $f\sim x$ or $f\sim y$ and follows from Lemma Lemma \ref{lem:time-equivalents} in the remaining cases. \\
We now show $V(f)=U(x_Ay)$. This follows from Lemma \ref{lem:repr_acts_bounded} if $f$ is bounded. Otherwise, consider the following exhaustive cases:\\
\textit{(Case 1)} $f$ is bounded from below, and not bounded from above. Since $u$ is bounded, there exists a sequence $\{x_n$, $n\geq 1\}$ of elements of $\sX$ such that $\{u(x_n)$, $n\geq 1\}$  converges to $\text{sup}\ u( f)$. For all $n\geq 1$, let $E_n^+\in\sB$ be the subset defined as the collection of all $(s,t)\in\sS\times\sT$ such that $u[f(s,t)]>u(x_n)$. Since $f$ is not bounded from above,  $\{E_n^+$, $n\geq 1\}$ is vanishing.  By Step 1, for all $n\geq 1$, there exist $\overline{x}_n,\underline{x}_n\in\sX$ such that ${\overline{x}_n}_{E_n^+}f\succsim f\succsim {\underline{x}_n}_{E_n^+}f$ and, therefore, ${\overline{x}_n}_{E_n^+}f\succsim x_Ay\succsim {\underline{x}_n}_{E_n^+}f$. Since the three acts are bounded, Lemma \ref{lem:repr_acts_bounded} yields
\[V({\overline{x}_n}_{E_n^+}f)\ \geq \ U(x_Ay)\ \geq\ V({\underline{x}_n}_{E_n^+}f).\]
By Step 2, taking limits gives $V(f)=U(x_Ay)$.\\
\textit{(Case 2)} $f$ is bounded from above, and not bounded from below. Consider then a sequence $\{y_n$, $n\geq 1\}$ of elements of $\sX$ such that $\{u(y_n)$, $n\geq 1\}$  converges to $\text{inf}\ u (f)$ and, for all $n\geq 1$, let  $E_n^-\in\sB$ be  the collection of all $(s,t)\in\sS\times\sT$ such that $u[f(s,t)]<u(y_n)$. Since $f$ is not bounded from below, $\{E_n^-$, $n\geq 1\}$ is vanishing. As in Case 1, we obtain $V(f)=U(x_Ay)$ again.\\
\noindent \textit{(Case 3)} $f$ is neither  bounded from below nor from above. Then, define $\{E_n^+$, $n\geq 1\}$ and $\{E_n^-$, $n\geq 1\}$ as in Cases 1 and 2. These are again vanishing sequences. For all $n\geq 1$, let $E_n=E_n^+\cup E_n^-$. Then, $\{E_n$, $n\geq 1\}$ is another vanishing sequence. Proceeding as in Cases 1 and 2, we obtain $V(f)=U(x_Ay)$ once more.

\noindent Now, consider $f,g\in\sF$. By the previous paragraphs, there exist $x,y,x',y'\in\sX$ and $A,B\in\sB_\sT$ such that $f\sim x_Ay$ and $g\sim x'_{B}y'$ with $V(f)=U(x_Ay)$ and $V(g)=U(x'_{B}y')$. Then,
\[f\ \succsim \ g\ \Longleftrightarrow\ x_Ay\ \succsim\ x'_By'\ \Longleftrightarrow\ U(x_Ay)\ \geq\ U(x'_By')\ \Longleftrightarrow\ V(f)\geq\ V(g),\]
where the second equivalence is by Proposition \ref{prop:rep_finite_act}.
\ethmprf

\section{Proof of Theorem \ref{main}}\label{AppD}
\setcounter{thm}{0} \setcounter{equation}{0}
\renewcommand{\theequation}{D\arabic{equation}}
\renewcommand{\thethm}{D\arabic{thm}}

We now come to the proof of Theorem \ref{main}. Proposition \ref{prop:rep_all_act} establishes the sufficiency of the axioms for the representation. Moreover, the uniqueness of $\lambda$ is implied by Proposition \ref{prop:discounting} while the uniqueness of $\mu$ and $u$ is implied by Proposition \ref{prop:rep_finite_act}.

\medskip\noindent Finally, as for the necessity of the axioms, suppose $(\lambda,u,\mu)$ provides a representation as in Theorem \ref{main}. Let $U$ and $V$ be the representing functionals for $\succsim_\sT$ and $\succsim$ as defined in Appendix \ref{AppC}. \textsc{$\sT$-Separability} follows from the remark that, for all disjoint $E_\sT,F_\sT\in\sB_\sT$, all $\bx\in\sX^\sT$ and all $x^*,x\in\sX$ with $x^*\succ_\sT x$, we have  $x^*_{E_\sT}x_{F_\sT}\bx\succsim_\sT  x_{E_\sT}x^*_{F_\sT}\bx$ if and only if $\eps_\lam(E_\sT)\geq \eps_\lam(F_\sT)$, where we assume $u(x^*)=1$ and $u(x)=0$ without loss of generality. As for \textsc{$\sT$-Measurability}, fix $\bx\in\sX^\sT$. Then, we have 
\[\{x\in\sX,\ x\succ_\sT \bx\}=u^{-1}(]\alp,+\infty[)\quad\text{and}\quad \{x\in\sX, \ \bx\succ_\sT x\}=u^{-1}(]-\infty,\alp[),\]
where $\alp=U(\bx)$. Then, \textsc{$\sT$-Measurability} follows from the measurability of $u$.  To show \textsc{Monotone Continuity}, suppose $f,g\in\sF$ are such that $f\succ g$, and consider $x\in\sX$ and a vanishing sequence $\{E_n$, $n\geq 1\}$ of subsets in $\sB$. By Step 2 of Proposition~\ref{prop:rep_all_act}, $\{V(x_{E_n}f)$, $n\geq 1\}$ and $\{V(x_{E_n}g)$, $n\geq 1\}$ converge respectively to $V(f)$ and $V(g)$. By the representation, we have $V(f)>V(g)$. Hence, there exists $N\geq 1$ such that $V(x_{E_N}f)>V(g)$ and $V(f)>V(x_{E_N}g)$. Then, still by the representation, we obtain $x_{E_N}f\succ g$ and $f\succ x_{E_N}g$.

 \Lemma{\label{lem1834}}
{Let $(\Omega,\sA,P)$ be a {\rm (}countably additive{\rm )} probability space. For all $E\in\sA$ and all measurable real-valued function $F$ on $\Omega$ such that $F(\omega)\geq 0$ for all $\omega\in\Omega$ and $F(\omega)>0$ for all $\omega\in E$, if $P(E)>0$, then $\int_\Omega F(\omega)dP(\omega)>0$. 
}

\bthmprf A proof is given for the sake of completeness. For all $n\geq 1$, let $E_n\in\sB$ be the subset of $E$ collecting all $\omega\in\Omega$ such that $F(\omega)\geq 1/n$. Then, the union of $\{E_n$, $n\geq 1\}$ is equal to $E$. By countable additivity, the limit of $\{P(E_n)$, $n\geq 1\}$ is equal to $P(E)$ so there exists $N\geq 1$ such that $P(E_N)>0$. We obtain
\[\int_\Omega F(\omega)dP(\omega)\ \geq \ \int_{E_N}F(\omega)dP(\omega)\ \geq\ \frac{1}{N}\cdot P(E_N)\ >\ 0,\]
where the first equality is by the (weak) monotonicity of the Lebesgue integral. 
 \ethmprf
 
\noindent Now, to show \textsc{$\sT$-Monotonicity},  let $\bx,\by\in\sX^\sT$ be such that  $\bx(t)\succsim_\sT \by(t)$ for all $t\in\sT$. Then, by the representation, we have $u(\bx(t))\geq u(\by(t))$ for all $t\in\sT$ and obtain $U(\bx)\geq U(\by)$ by the (weak) monotonicity of the Lebesgue integral. The representation further yields $\bx\succsim_\sT \by$. If additionally,  $\bx(t)\succ_\sT \by(t)$ for all $t$ in some non-null subset $E_\sT\in\sB_\sT$, then  $u(\bx(t))> u(\by(t))$ for all $t\in E_\sT$. As $E_\sT$ is non-null, we have $(\mu\times\eps_\lam)(\sS\times E_\sT)>0$. Lemma~\ref{lem1834} yields $U(\bx)> U(\by)$, and we obtain  $\bx\succ_\sT \by$ by the representation. As for \textsc{Dominance}, note that, for all $f\in\sF$, we have by the Fubini theorem
\[V(f)\quad =\quad \int_\sS\int_\sT u[f(s,t)]d\eps_\lam(t)d\mu(s).\]
Then, \textsc{Dominance} follows from an argument similar to that yielding \textsc{$\sT$-Monotonicity}. Finally, \textsc{Stationarity} follows from Lemma \ref{lem1835}. 

 \Lemma{\label{lem1835}}
{For all $t\in\sT$ and $f,h\in\sF$, 
\[V(h_tf)\ =\ \int_\sS\int_\sT\bone_{[0,t)}\cdot u[h(s,t')]d\eps_\lam(t')d\mu(s)\ +\ e^{-\lam t}\cdot V(f).\] 
}

\bthmprf By standard arguments, Formula (\ref{cauchy}) extends into the following one: For all $t\in\sT$  and $\bx\in\sX^\sT$
\[\int_\sT\bone_{[t,+\infty[}\cdot u[\bx(t'-t)]d\eps_\lam(t')\ =\ e^{-\lam t}\cdot U(\bx).\]
Then, we have for all $t\in\sT$ and $\bx,\bz\in\sX^\sT$
\[U(\bz_t\bx)\ =\ \int_\sT\bone_{[0,t)}\cdot u[\bz(t')]d\eps_\lam(t')\ +\ e^{-\lam t}\cdot U(\bx).\] 
For all $t\in\sT$ and $f,h\in\sF$, applying the previous formula to $f(s,\cdot)$ and $h(s,\cdot)$ for all $s\in\sS$ and integrating on $\sS$ yields the result.\ethmprf

 {\footnotesize 
\bibliographystyle{elsart-harv}
\bibliography{Biblio2}
 \end{document}